\documentclass[12pt,english,spanish]{article}
\usepackage[T1]{fontenc}
\usepackage[utf8]{inputenc}
\usepackage[a4paper]{geometry}
\usepackage{babel}
\addto\shorthandsspanish{\spanishdeactivate{~<>}}

\usepackage{float}
\usepackage{graphicx}
\usepackage{setspace}
\usepackage[unicode=true,pdfusetitle,
 bookmarks=true,bookmarksnumbered=false,bookmarksopen=false,
 breaklinks=false,pdfborder={0 0 0},backref=false,colorlinks=false]
 {hyperref}

\makeatletter

\providecommand{\tabularnewline}{\\}

\newenvironment{lyxcode}
{\par\begin{list}{}{
\setlength{\rightmargin}{\leftmargin}
\setlength{\listparindent}{0pt}% needed for AMS classes
\raggedright
\setlength{\itemsep}{0pt}
\setlength{\parsep}{0pt}
\normalfont\ttfamily}%
 \item[]}
{\end{list}}

\@ifundefined{date}{}{\date{}}
\usepackage[T1]{fontenc}
\usepackage{lmodern}

\makeatother

\usepackage{listings}
\addto\captionsenglish{}
\addto\captionsspanish{}

\begin{document}

\title{Análisis econométrico de series temporales en Gretl: La Ley de Okun\thanks{Licenciado bajo \protect\href{https://creativecommons.org/licenses/by-nc-sa/4.0/legalcode}{Creative Commons Attribution-NonCommercial-ShareAlike 4.0.}}}

\author{Eduardo Calvo del Río\thanks{Email: \protect\href{mailto:eduardo.calvo.rio@alumnos.uva.es}{eduardo.calvo.rio@alumnos.uva.es}}}
\maketitle
\begin{abstract}
Gretl es un software econométrico de código abierto que se presenta
como una alternativa muy potente al software comercial. El objeto
de este trabajo es demostrar la facilidad y versatilidad de este programa
en el análisis econométrico de series temporales. Para ello se estudiará
la Ley de Okun que es aquella relación existente entre la variación
de la tasa de crecimiento y la tasa de paro en una economía.\textsc{}
\end{abstract}

\section{Introducción}

\subsection{Gretl}

Gretl (Gnu Regression, Econometrics and Time-series Library) es un
paquete de software para análisis econométrico escrito en el lenguaje
de programación C y distribuido bajo la licencia GPL de la Free Software
Foundation. 

El código de Gretl se deriva del programa ESL (Econometrics Software
Library) escrito por el profesor Ramu Ramanathan de la Universidad
de California \cite{ramanathan1998introductory}. La distribución
de Gretl como software libre ha permitido que multitud de usuarios
participen en su desarrollo, depuración del código, traducción y divulgación
por ámbitos académicos y empresariales.

Gretl es un software especialmente útil en la enseñanza de la Econometría
porque incluye los ejemplos y ejercicios de los principales manuales
de la materia.\footnote{La lista incluye los manuales de \cite{wooldridge2006introduccion,gujarati2010econometria,greene1999analisis},
entre otros.}

Entre las ventajas de Gretl respecto de otras soluciones comerciales
está su arquitectura multiplataforma que permite instalarlo en distintos
sistemas operativos y la dualidad de interfaz que hace posible su
ejecución en modo gráfico pero también en modo de línea de comandos
(CLI). Además, Gretl ofrece toda una serie de modelos econométricos
(MCO, MC2E, ARIMA, GARCH, GMM, VI...) que lo dotan de un alto grado
de funcionalidad.

\section{Revisión del estado actual del tema}

\subsection{La Ley de Okun}

El ejemplo utilizado para mostrar la funcionalidad y flexibilidad
de Gretl es la Ley de Okun. Esta es una relación que se ha venido
dando entre la tasa de crecimiento de una economía y su tasa de desempleo. 

Esta Ley ha sido objeto de numerosos trabajos y se encuentra en la
práctica totalidad de manuales de Macroeconomía por su amplia difusión
y aceptación por todas las corrientes del pensamiento económico. La
actual situación laboral en España es lo que me ha llevado a elegir
esta regularidad como objeto de estudio y análisis.

Arthur Okun, economista y miembro del Consejo de Asesores Económicos
del presidente John F. Kennedy, reconoció esta regularidad en 1962
con datos trimestrales de los Estados Unidos. 

Okun tomó tres vías diferentes \cite{okun1963potential} para estimar
la relación entre la tasa de paro y de crecimiento de la producción.

\subsubsection{Ley de Okun I}

En primer lugar, Okun estimó un modelo que relacionaba las variaciones
de la tasa de paro en dos períodos consecutivos y la tasa de crecimiento
de la producción

\begin{equation}
u_{t}-u_{t-1}=\beta_{0}-\beta_{Y}g_{Yt}\label{eq:okuni1}
\end{equation}

Donde $u_{t}$ es la tasa de paro y $g_{Yt}$ la tasa de crecimiento
de la producción en el periodo $t$. Si la variación de la tasa de
paro es nula, el valor de $g_{Yt}$ representa la tasa de crecimiento
natural de la producción en el periodo $t$. Lo denotamos $\overline{g}_{Yt}$
\cite{blanchard2006macroeconomia}.

\begin{equation}
\beta_{0}-\beta_{Y}g_{Yt}=0;\;\overline{g}_{Yt}=\frac{\beta_{0}}{\beta_{Y}}
\end{equation}

Sustituyendo $\beta_{0}$ en la ecuación (\ref{eq:okuni1}), se obtiene
que la variación de la tasa de paro es proporcional a la diferencia
entre la tasa de crecimiento real y natural de la producción.

\begin{equation}
u_{t}-u_{t-1}=-\beta_{Y}\left(g_{Yt}-\overline{g}_{Yt}\right)
\end{equation}

\subsubsection{Ley de Okun II}

Un enfoque alternativo llevó a Okun a plantear un modelo en el que
relacionaba la tasa de paro con la diferencia entre la producción
potencial y real.

\begin{equation}
u_{t}=\lambda_{0}+\lambda_{y}\frac{Y_{t}^{p}-Y_{t}}{Y_{t}^{p}}
\end{equation}

El problema de este modelo es que la producción potencial $Y_{t}^{p}$
es una variable desconocida a la que Okun le asigna un valor de forma
subjetiva y arbitraria aunque posteriormente se han utilizado procedimientos
estadísticos como el filtro H-P para estimar su valor \cite{belmonte2004formulaciones}.

\subsubsection{Ley de Okun III}

La última versión del modelo que Okun propone no parte de la tasa
de paro sino de la tasa de ocupación potencial $e_{t}^{p}$ y real
$e_{t}$. Okun estimó que la relación entre ambas era proporcional
al cociente entre la producción potencial y real. Es decir, que la
eficiencia del trabajo es proporcional a la eficiencia de la producción.

\begin{equation}
\frac{e_{t}}{e_{t}^{p}}=\left(\frac{Y_{t}}{Y_{t}^{p}}\right)^{\sigma_{Y}}
\end{equation}

Donde $\sigma_{Y}$ es la elasticidad del empleo respecto de la producción
en valor absoluto.

A pesar de las diferencias conceptuales existentes entre las tres
formulaciones, los valores estimados son muy similares \cite{okun1963potential}.
$\beta_{Y}=0.3$, $\lambda_{y}=0$, 36 y $\sigma_{Y}=0.35$.

\subsubsection{La Ley de Okun en este trabajo\label{sub:okun generalizado}}

Para el análisis econométrico de la Ley de Okun se va a partir de
una formulación más reciente de la primera versión \cite{belmonte2004formulaciones}.
Este modelo es una generalización que incluye retardos tanto en la
variable independiente como en la dependiente.

\begin{equation}
\Delta u_{t}=-\beta_{0}+\sum_{1}^{P}{\beta_{up}\Delta u_{t-p}}+\sum_{0}^{Q}{\beta_{yp}\Delta\ln{\left(y_{t-q}\right)}}+\varepsilon_{t}
\end{equation}

\section{Metodología}

Es importante tener claros todos los conceptos que intervienen en
el análisis econométrico de series temporales relacionados con esta
investigación. En las siguientes secciones se van a repasar y clarificar
todos estos conocimientos previos que permitirán realizar un análisis
econométrico adecuado a partir de series cronológicas.

\subsection{Esquema autorregresivo de primer orden}

Parece conveniente empezar este estudio de la series temporales a
partir de un proceso estocástico muy habitual. Se trata del esquema
autorregresivo de primer orden, también denominado $AR(1)$, para
abreviar. Como explican \cite{gujarati2010econometria}, en estos
procesos, la variable dependiente es una variable estocástica formada
por su valor en el momento anterior y por un término de error. Además,
puede aparecer un término constante u ordenada en el origen.

\begin{eqnarray}
Y_{t} & = & \alpha_{0}+\varphi Y_{t-1}+\varepsilon_{t}
\end{eqnarray}

La variable retardada está ponderada por un valor $\varphi$ en el
intervalo $\left[-1,\,1\right]$. Igualmente, se puede generalizar
el proceso autorregresivo para órdenes superiores a uno, añadiendo
los retardos pertinentes. Por lo tanto se obtiene un esquema $AR(p)$:

\begin{eqnarray}
Y_{t} & = & \alpha_{0}+\sum_{i=1}^{p}\varphi_{i}Y_{t-i}+\varepsilon_{t}
\end{eqnarray}

El sistema de scripts de Gretl permite generar simulaciones computacionales
de cualquier fenómeno estadístico. Se ha elaborado el siguiente script\footnote{La ejecución de un script se realiza con el comando \texttt{run {[}archivo.inp{]}}}
para dibujar gráficos de modelos $AR(1)$ con distintos valores de
$\varphi$.

\inputencoding{latin9}\begin{lstlisting}[basicstyle={\scriptsize\ttfamily}]
# fija el tamaño muestral y lo clasifica como temporal
nulldata 100
setobs 1 1 --time-series
genr time

# semilla de números aleatorios
set seed 7777777

# parámetros fijos
scalar phi = .5

# generación de series
series y = uniform()
series e = normal()

# modelo AR(1)
series y = phi * y(-1) + e

# representación gráfica
gnuplot y --with-lines --time-series
corrgm y 50
 
\end{lstlisting}
\inputencoding{utf8}

Que genera los gráficos siguientes

\begin{figure}[H]
\begin{centering}
\includegraphics{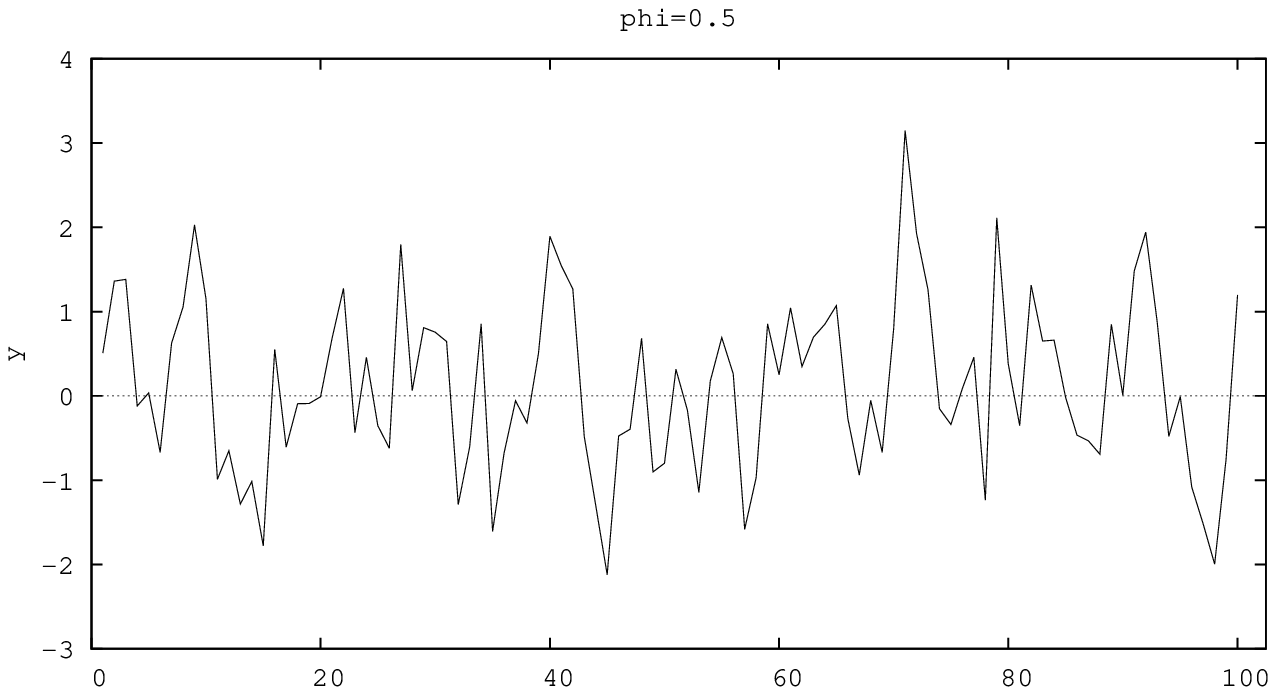}\\
\includegraphics{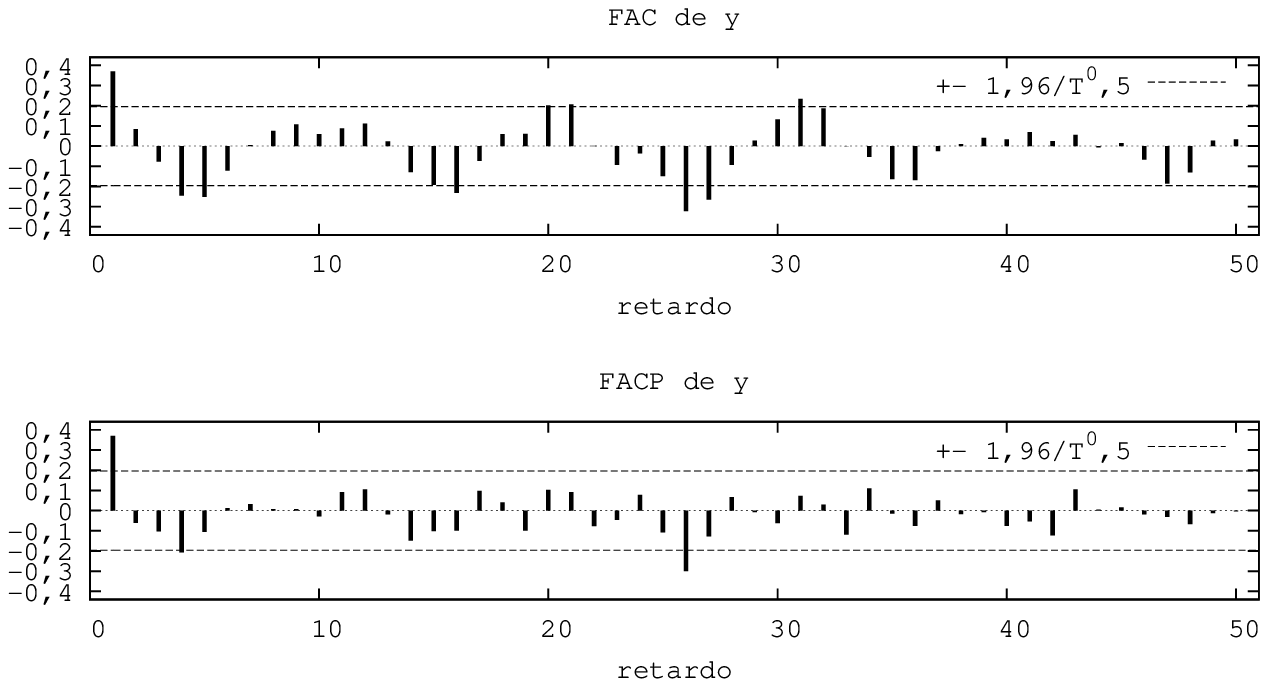}
\par\end{centering}

\centering{}\caption{$AR(1),\,\varphi=0.5$ }
\end{figure}

\begin{figure}[H]
\begin{centering}
\includegraphics{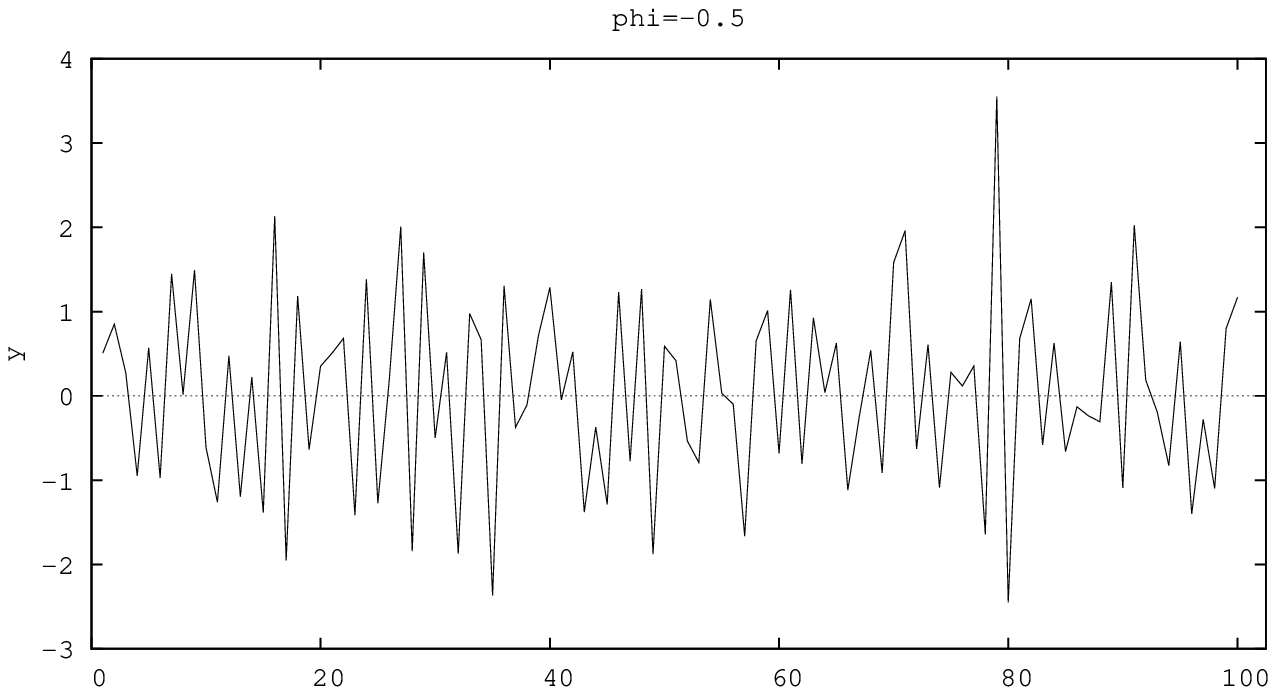}\\
\includegraphics{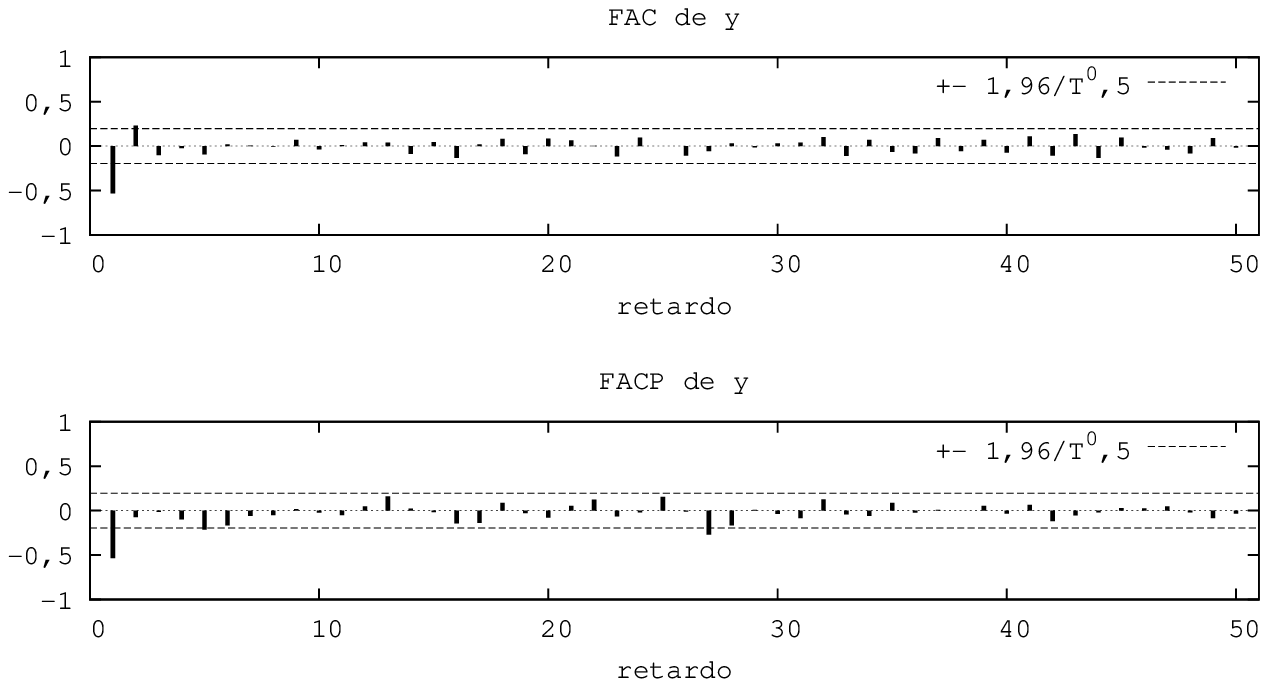}
\par\end{centering}

\centering{}\caption{$AR(1),\,\varphi=-0.5$ }
\end{figure}

\subsection{Caminata aleatoria}

Es una situación particular de esquema autorregresivo. Concretamente,
se produce cuando $\varphi=1$ y por tanto $Y_{t}=Y_{t-1}+\varepsilon_{t}$.
Para ver mejor sus propiedades, es conveniente desarrollar su ecuación
fundamental desde la primera observación.

\[
Y_{1}=Y_{0}+\varepsilon_{1}
\]

\[
Y_{2}=Y_{1}+\varepsilon_{2}=Y_{0}+\varepsilon_{1}+\varepsilon_{2}
\]

\[
\vdots
\]

\begin{equation}
Y_{t}=Y_{0}+\sum_{i=1}^{t}\varepsilon_{i}
\end{equation}

Con cada nueva iteración del proceso, el error aumenta, en valor absoluto.
Esto se refleja en la varianza de la caminata aleatoria que es $VAR\left(Y_{t}\right)=\sigma^{2}t$
por lo que esta serie no estará acotada. Reconfigurando el script
del $AR(1)$ se puede generar un gráfico que ilustra las propiedades
de este proceso estocástico.

\begin{figure}[H]
\begin{centering}
\includegraphics{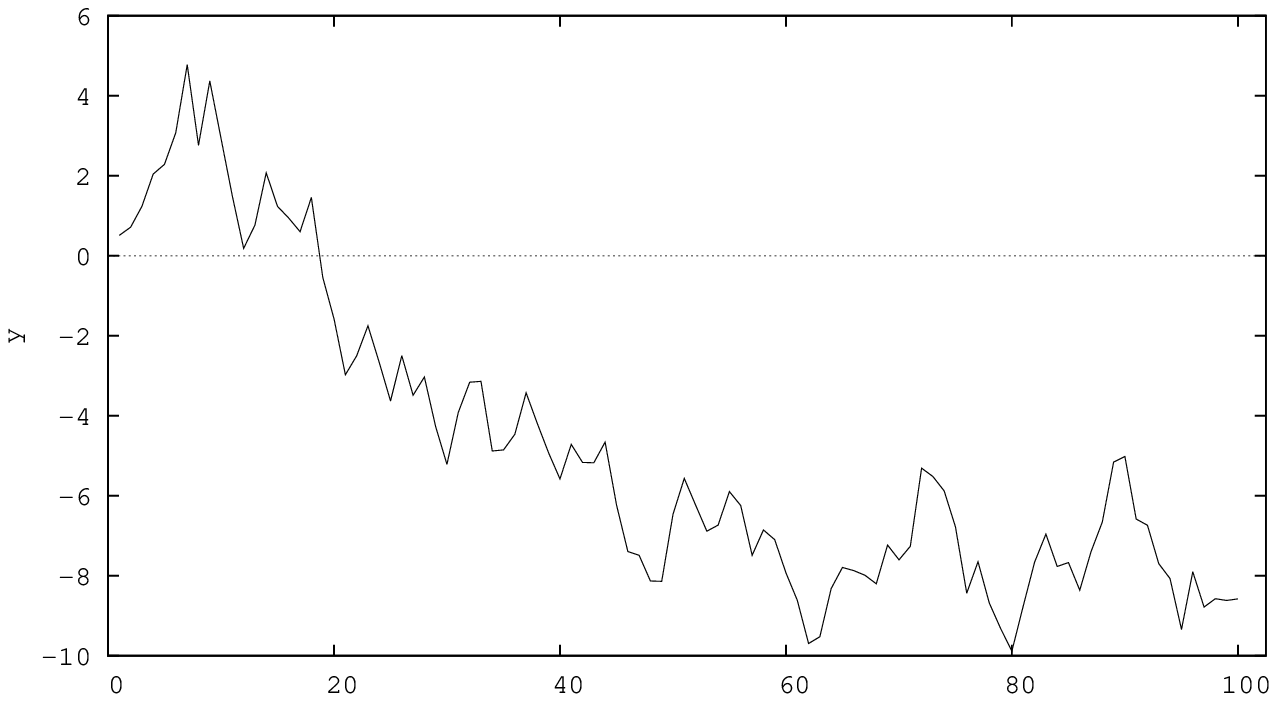}\\
\includegraphics{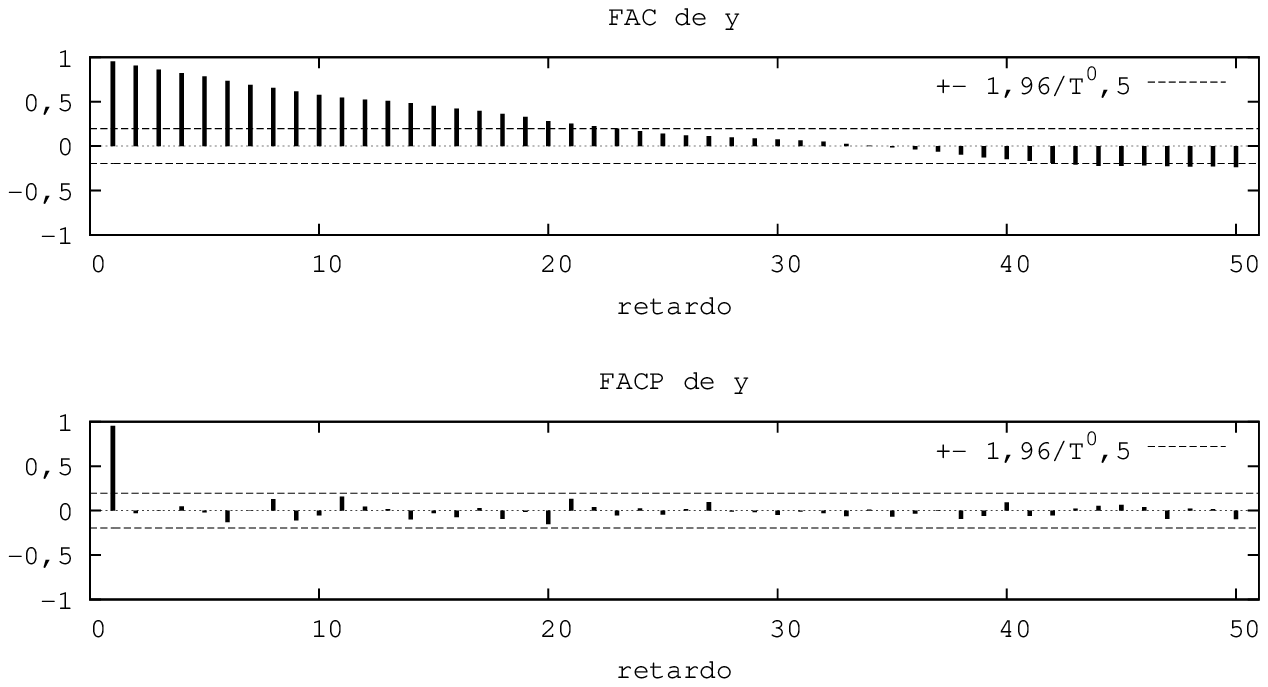}
\par\end{centering}

\centering{}\caption{Caminata aleatoria sin deriva}
\end{figure}

\begin{figure}[H]
\begin{centering}
\includegraphics{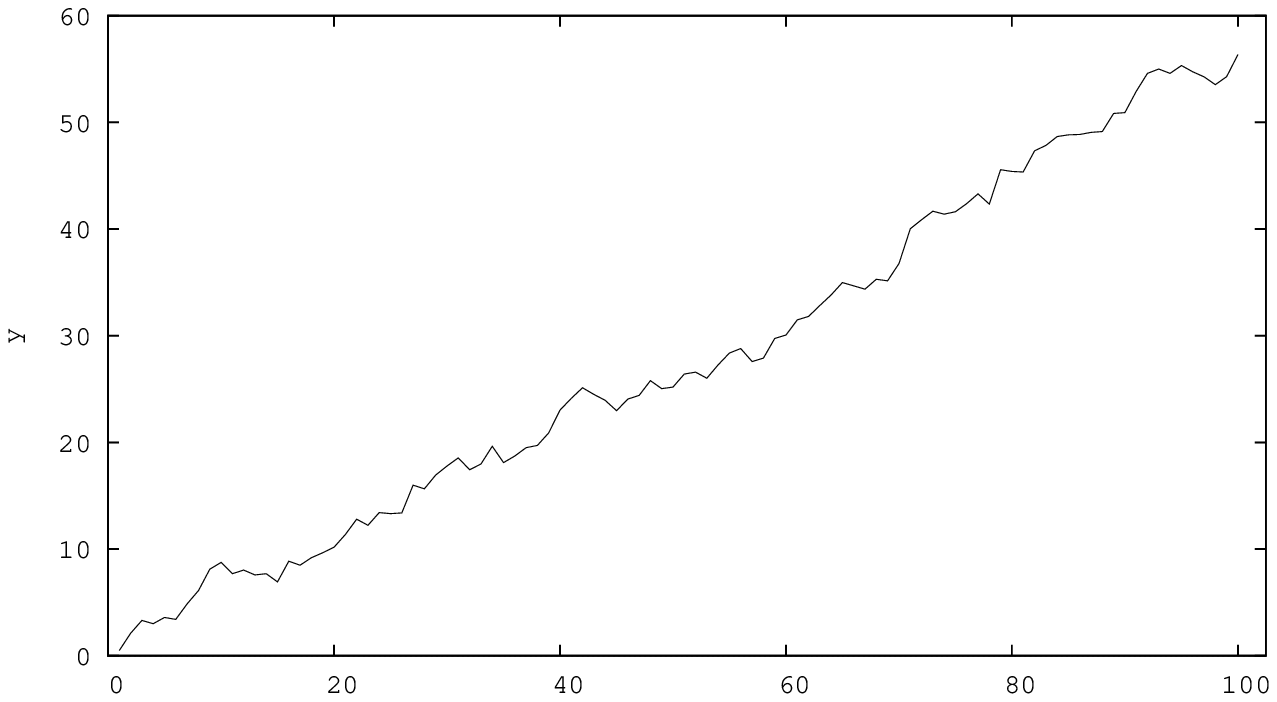}\\
\includegraphics{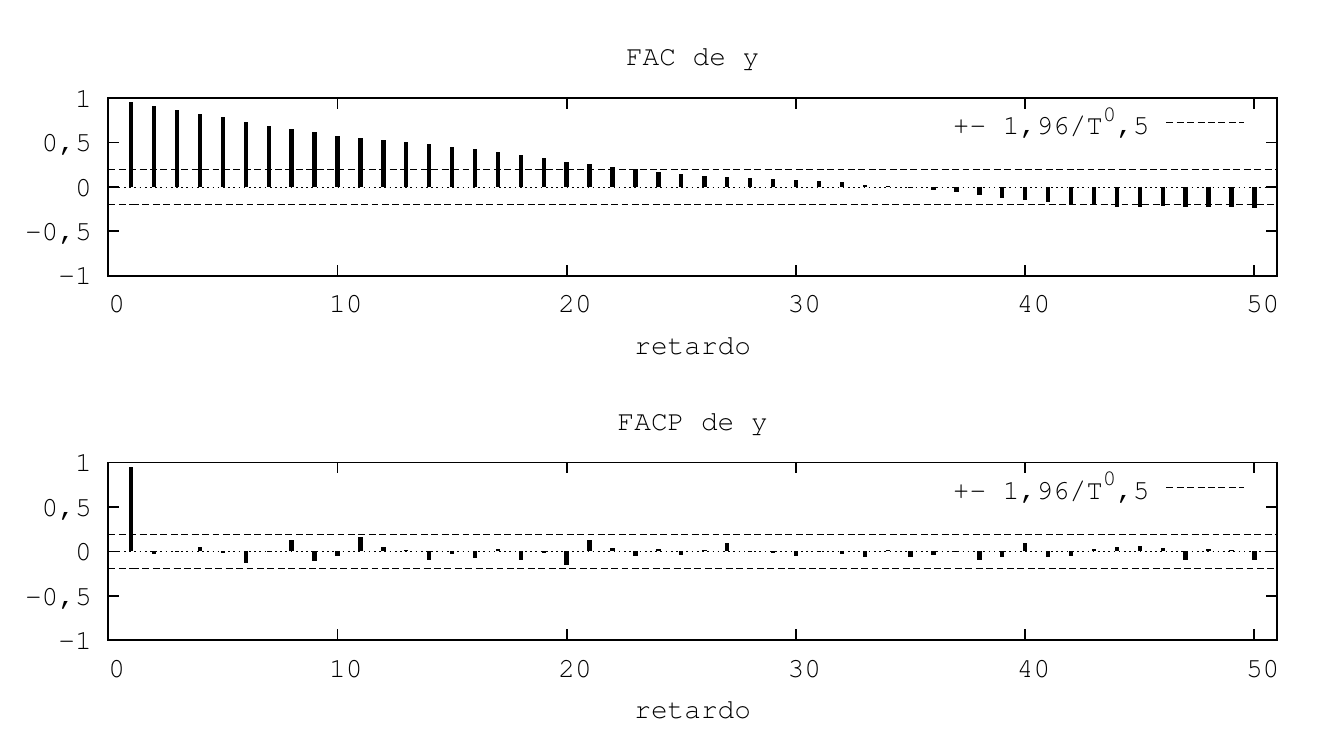}
\par\end{centering}

\centering{}\caption{Caminata aleatoria con deriva. $\alpha=0.5$}
\label{caminataconderiva}
\end{figure}

Ahora bien, hay otro tipo de caminata aleatoria en la que se incluye
un término constante. Se denomina caminata aleatoria con deriva y
sus propiedades difieren de la anterior en que su varianza sí está
acotada aunque su media se mueve en proporción al término de la deriva,
como se muestra en la figura \ref{caminataconderiva}.

\begin{equation}
Y_{t}=\alpha+Y_{t-1}+\varepsilon_{t}
\end{equation}

\subsection{Ruido blanco}

Un ruido blanco es un proceso estocástico normal, incorrelacionado
y homoscedástico. Es el pilar fundamental sobre el que se asienta
el modelo clásico de econometría. El término de error en un modelo
clásico es, por definición, una variable desconocida pero que se distribuye
de una forma particular, que conocemos de antemano. Es una normal
de varianza constante. 

\begin{equation}
\varepsilon\sim N\left(0,\sigma^{2}I\right)
\end{equation}

Diversos fenómenos de autocorrelación y heteroscedasticidad incumplen
las hipótesis clásicas. Esto debe ser tenido en cuenta a la hora de
formular un modelo econométrico para no asignarle ciertas cualidades
estadísticas que en realidad no tenga.

Para acompañar la explicación, también se ha programado el siguiente
script que genera ruido blanco.

\inputencoding{latin9}\begin{lstlisting}[basicstyle={\scriptsize\ttfamily}]
# fija el tamaño muestral y lo clasifica como temporal
nulldata 100
setobs 1 1 --time-series
genr time

# semilla de números aleatorios
set seed 7777777

series e = normal()

# representación gráfica
gnuplot e --with-lines --time-series
corrgm e 50
 
\end{lstlisting}
\inputencoding{utf8}

\begin{figure}[H]
\begin{centering}
\includegraphics{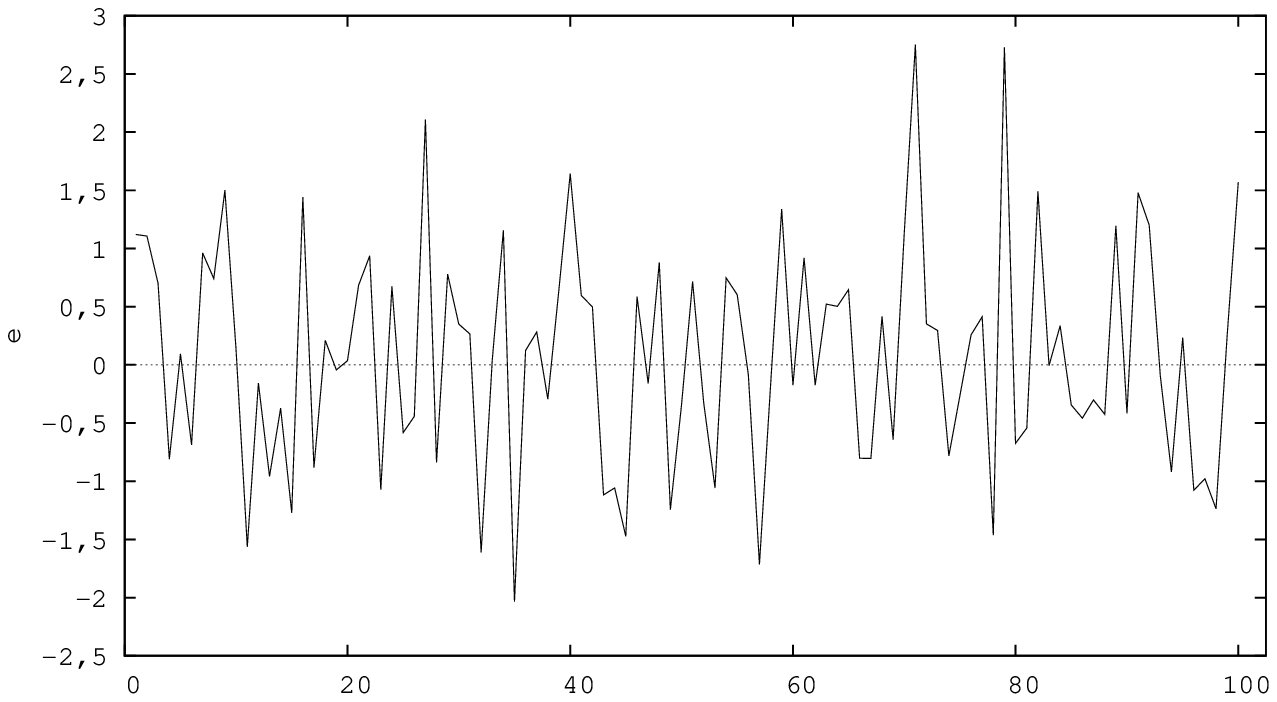}\\
\includegraphics{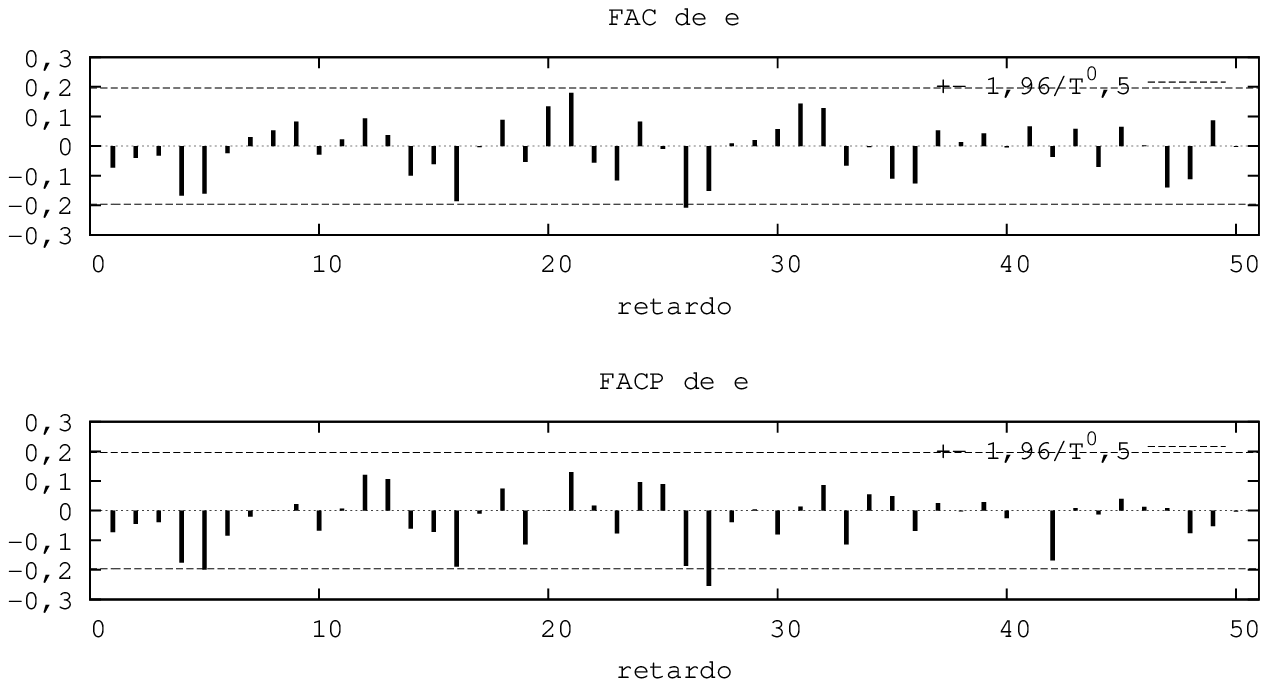}
\par\end{centering}

\centering{}\caption{Ruido blanco }
\end{figure}

\subsection{Ergodicidad}

El concepto de ergodicidad se refiere a la convergencia probabilística
de los momentos muestrales en los poblacionales cuando $s\rightarrow\infty$,
es decir, cuando la serie se aleja del momento $t$. Esto supone que
la función de autocorrelación observada en los correlogramas tiende
a cero conforme crece el desfase entre los momentos $t$ y $s$, es
decir, $plim\left(\rho_{s}\right)=0.$

\subsection{Estacionariedad}

\label{sec:section}En el análisis econométrico de series temporales
es de vital importancia determinar si el proceso estocástico con el
que se trabaja es estacionario o integrado de orden $d$, pues de
esto va a depender la validez asintótica de toda la modelización que
se lleve a cabo. Para clasificar siguiendo este procedimiento a una
serie estocástica, hay que fijarse en los momentos de las distribuciones
objeto de estudio, ya que es imposible conocer con certeza la función
de distribución conjunta del proceso estocástico en cada periodo.

Formalmente se define un proceso estocástico como un conjunto de variables
aleatorias que se suceden en el tiempo.

\[
Y=\left\{ Y_{1},\,Y_{2},\,\dots,\,Y_{T}\right\} 
\]

Cuyos momentos tienen la forma:

\[
E\left(Y_{t}\right)=\mu_{t}
\]

\[
var\left(Y_{t}\right)=\sigma_{t}^{2}
\]

\selectlanguage{english}%
\[
cov\left(Y_{t},Y_{t+s}\right)=\sigma_{t,t+s}
\]

\selectlanguage{spanish}%
Se puede afirmar que un proceso estocástico es estacionario en sentido
débil si se cumple que la esperanza y varianza son constantes a lo
largo del tiempo y que la covarianza entre variables en distintos
momentos de tiempo solo depende del desfase entre ellas.

\[
\mu_{t}=\mu
\]

\[
\sigma_{t}^{2}=\sigma^{2}
\]

\[
\sigma_{t,t+s}=\sigma_{t,t-s}
\]

Pero generalmente las variables económicas no son estacionarias ya
que presentan algún tipo de tendencia que provoca que la media y/o
varianza no sean constantes en el tiempo. Una serie económica es integrada
de orden $d$ si se puede transformar en una serie estacionaria aplicando
diferencias de orden $d$, es decir, tomando primeras diferencias,
$d$ veces. En concreto, si $X_{t}$ es $I(1)$, entonces $Y_{t}=X_{t}-X_{t-1}$
es una serie estacionaria.

\subsubsection{Test de Dickey-Fuller}

El test de Dickey y Fuller parte de un proceso autorregresivo de orden
uno. Estos procesos estocásticos se convierten en modelos de caminata
aleatoria si tienen raíz unitaria, es decir si $\varphi=1$. Por lo
tanto el contraste DF hace una regresión auxiliar de $Y_{t}$ sobre
su valor rezagado un periodo para comprobar la anterior igualdad. 

\begin{equation}
Y_{t}=\varphi Y_{t-1}+\varepsilon_{t}\label{eq:DF}
\end{equation}

La ecuación \ref{eq:DF} es la expresión básica del test DF. A esta
se le puede añadir un término constante, una variable de tendencia
determinista e incluso de tendencia cuadrática.

Las pruebas realizadas mediante experimentos de Monte Carlo han demostrado
que este test presenta un sesgo muy marcado si existe verdaderamente
la raíz unitaria\cite{gujarati2010econometria}. La solución planteada
es restar un retardo a ambos lados de la ecuación. 

\[
Y_{t}-Y_{t-1}=\varphi Y_{t-1}-Y_{t-1}+\varepsilon_{t}
\]

\begin{equation}
\Delta Y_{t}=\underbrace{\left(\varphi-1\right)}_{\delta}Y_{t-1}+\varepsilon_{t}
\end{equation}

Además, esto provoca un cambio de variable que se pretende contrastar,
que ahora será $\delta=\varphi-1$. Consecuentemente, la nueva hipótesis
nula será $H_{0}:\delta=0$ frente a la alternativa $H_{1}:\delta<0$.
El problema de este contraste radica en que la distribución que sigue
el estadístico no es la típica t de Student sino una distribución
cuyos valores recogen \cite{dickey1979distribution} y posteriormente
completa \cite{mackinnon1996numerical}\footnote{Gretl utiliza los valores de Davidson y MacKinnon. Las tablas completas
pueden consultarse en el artículo citado.}.

\subsubsection{Test Aumentado de Dickey-Fuller\label{sub:adf}}

El test DF no tiene en cuenta la existencia de autocorrelación, situación
muy habitual en series temporales. El planteamiento que soluciona
este problema es el test ampliado de Dickey y Fuller. En la práctica
es el test que se utiliza para contrastar la estacionariedad. 

Analíticamente el contraste ADF es

\begin{equation}
\Delta Y_{t}=\beta_{0}+\delta Y_{t-1}+\sum_{i=1}^{q}{\gamma_{i}\Delta Y_{t-1}}+\varepsilon_{t}
\end{equation}

El test ADF tiene una variante que se utiliza cuando la serie presenta
una tendencia temporal. En ese caso habría que añadir la variable
tiempo como un regresor más de la ecuación:

\begin{equation}
\Delta Y_{t}=\beta_{0}+\alpha t+\delta Y_{t-1}+\sum_{i=1}^{q}{\gamma_{i}\Delta Y_{t-1}}+\varepsilon_{t}
\end{equation}
En este test se añaden retardos de la primera diferencia hasta eliminar
la autocorrelación. El número de rezagos tiene que ser el mínimo posible
para que el test no pierda potencia. Gretl, a diferencia de otros
paquetes econométricos comerciales, selecciona automáticamente este
valor, siguiendo un criterio de información de Akaike modificado \cite{ng2001lag}.
Concretamente, el comando de Gretl para realizar un test de raíces
unitarias es el siguiente.

\inputencoding{latin9}\begin{lstlisting}[basicstyle={\scriptsize\ttfamily}]
adf [serie] [opciones] 
\end{lstlisting}
\inputencoding{utf8}

Cuyas principales opciones se explican a continuación.

\inputencoding{latin9}\begin{lstlisting}[basicstyle={\scriptsize\ttfamily},tabsize=4]
--c 							con constante
--ct 							con constante y tendencia
--difference 					usa la primera diferencia de la variable
--test-down=[criterio] 			selección automática del número de rezagos
\end{lstlisting}
\inputencoding{utf8}

\subsection{Cointegración}

Puede parecer que la existencia de raíces unitarias no es ningún obstáculo
ya que se puede resolver fácilmente aplicando diferencias, pero lo
que esto produce es un cambio en la especificación del modelo que
afecta a su interpretación. Lo que verdaderamente se estima en el
caso de que no haya estacionariedad no es una variable sino la variación
que esta ha sufrido en el último periodo, por lo que no es correcto
utilizar este enfoque para analizar procesos a largo plazo.

La Teoría Económica no puede estar limitada a hacer estudios cortoplacistas
aunque sabemos que la mayor parte de variables e indicadores económicos
presentan un problema provocado por la ausencia de estacionariedad. 

Afortunadamente, Engle y Granger encontraron una forma de superar
este obstáculo. Su gran aportación fue recompensada con el Premio
Nobel de Economía 2003.

Su concepto de cointegración explica que entre dos variables con el
mismo orden de integración puede existir una combinación lineal estacionaria
\cite{engle1987cointegration}. 

Sea un vector de variables $X_{t}=\left(X_{1},\,X_{2},\,\dots,\,X_{k}\right)$
en el que cada variable $X_{i}\sim I(d).$ Generalmente, las combinaciones
lineales entre variables serán también $I(d)$. Sin embargo, es posible
que las series estén cointegradas. $X_{t}\sim CI(d,\,b)$ si existe
un vector $\alpha$ no nulo tal que $Y_{t}=\alpha'X_{t}\sim I(d-b),\,b>0$.
El caso más interesante es aquel en que $d=b=1$ así que las variables
son integradas de orden 1 pero existe una combinación lineal suya
que es estacionaria. Esto sucede porque los efectos a largo plazo
de ambas variables se anulan mutuamente para lograr el equilibrio. 

Engle y Granger siguen distintos procedimientos para determinar la
cointegración de series temporales pero recomiendan utilizar una prueba
que parte del test ADF. En este caso:

\begin{equation}
\Delta u_{t}=\beta_{0}+\delta u_{t-1}+\sum_{i=1}^{q}{\gamma_{i}\Delta u_{t-1}}+\varepsilon_{t}
\end{equation}

Donde $u_{t}$ es el residuo de la regresión de cointegración entre
las variables del vector $X_{t}$ y la hipótesis nula implica ausencia
de cointegración $H_{0}:\delta=0$.

Sin embargo, en Gretl no es necesario realizar de forma manual las
sucesivas etapas que forman el test. Existe, al contrario que en otros
paquetes econométricos, un comando que ejecuta directamente todos
los pasos del test de Engle y Granger.\inputencoding{latin9}
\begin{lstlisting}[basicstyle={\scriptsize\ttfamily}]
coint [orden máximo de rezagos] [series] [opciones] 
\end{lstlisting}
\inputencoding{utf8}

Cuyas principales opciones son:

\inputencoding{latin9}\begin{lstlisting}[basicstyle={\scriptsize\ttfamily},tabsize=4]
--c 							con constante
--ct 							con constante y tendencia
--skip-df						omite un test ADF previo
--test-down						selección automática del número de rezagos
\end{lstlisting}
\inputencoding{utf8}

\subsection{Autocorrelación}

Aunque se esté usando el contraste de Durbin-Watson para detectar
la autocorrelación en las regresiones auxiliares de los test de estacionariedad,
hay que advertir que, en principio, este test se ciñe al contraste
de autocorrelación en forma de proceso $AR(1)$. 

Sería más completo utilizar también un contraste que superara esta
restricción para analizar el modelo que finalmente se plantee. Gretl
permite obtener el test de autocorrelación de Breusch-Godfrey, cuya
hipótesis alternativa es $H_{1}:\,AR(p)\:\acute{o}\:MA(q)$. El comando
que ejecuta este contraste es el siguiente:

\inputencoding{latin9}\begin{lstlisting}[basicstyle={\scriptsize\ttfamily}]
modtest [opciones] [orden]
\end{lstlisting}
\inputencoding{utf8}

Además, este comando da la posibilidad de calcular otros contrastes
muy importantes para el análisis econométrico. Se enumeran las principales
opciones a continuación.

\inputencoding{latin9}\begin{lstlisting}[basicstyle={\scriptsize\ttfamily}]
--normality 			normalidad de los residuos
--logs 				no-linealidad logarítmica
--squares 			no-linealidad cuadrática
--autocorr 			autocorrelación
--white 			test de heteroscedasticidad de White
--breusch-pagan 		test de heteroscedasticidad de Breusch-Pagan
\end{lstlisting}
\inputencoding{utf8}

\section{Resultados}

\subsection{Series temporales\label{sub:Series-temporales}}

Se parte de unos datos\footnote{Todos los datos, gráficos y códigos de este trabajo están disponibles
en mi repositorio git en \href{https://github.com/edkalrio/tfg}{https://github.com/edkalrio/tfg}} de desempleo y PIB de España en el periodo 1980-2012 obtenidos de
la base de datos del banco mundial.

Para evitar que este trabajo quede obsoleto ante un cambio de interfaz
en Gretl, se ha optado por presentar las instrucciones en forma de
comandos. Estos pueden introducirse en la consola de Gretl, a la que
se accede pulsando \texttt{c} en la pantalla principal de Gretl o
directamente desde la terminal del sistema con el comando \texttt{gretlcli.}

El primer paso es importar los datos con los que se va a trabajar.
Los comandos que le siguen, sirven para clasificar las series como
temporales y ordenarlas cronológicamente. Por último, es útil generar
una variable temporal que puede servir para hacer regresiones auxiliares
o representaciones gráficas. Como se aprecia, la sintaxis de Gretl
no es especialmente difícil. Usar comandos en vez de recurrir a la
interfaz gráfica permite trabajar con mayor velocidad.

\inputencoding{latin9}\begin{lstlisting}[basicstyle={\scriptsize\ttfamily}]
? open ~/tfg.xls
? setobs 1 1980 --time-series
? genr time
\end{lstlisting}
\inputencoding{utf8}

En el estudio de la estacionariedad, es importante obtener la representación
gráfica de las series temporales. Gretl ofrece la posibilidad de representar
gráficos múltiples para agrupar distintas series. En este caso, dibujar
la serie $y$ así como su primera diferencia.

\inputencoding{latin9}\begin{lstlisting}[basicstyle={\scriptsize\ttfamily}]
? scatters y d_y
\end{lstlisting}
\inputencoding{utf8}

\begin{figure}[H]
\centering{}\includegraphics{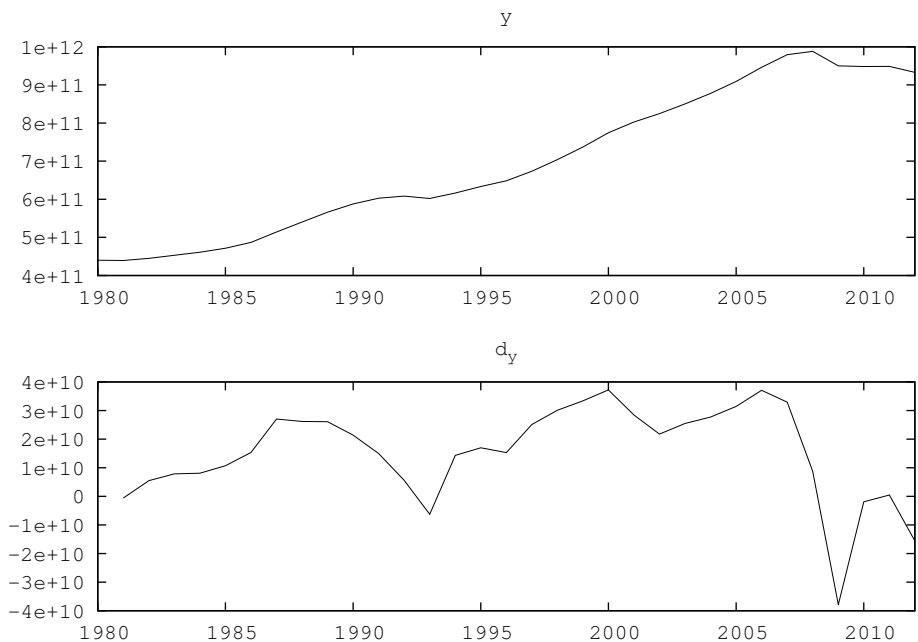}\caption{variable de la producción a nivel y primera diferencia}
\end{figure}

Se aprecia que la serie $y$ tiene tenencia lineal. Para cerciorarse,
solo hay realizar una regresión MCO auxiliar contra la variable temporal.
\inputencoding{latin9}\begin{lstlisting}[basicstyle={\scriptsize\ttfamily}]
? ols y const time
\end{lstlisting}
\inputencoding{utf8}

\inputencoding{latin9}\begin{lstlisting}[basicstyle={\scriptsize\ttfamily}]
Modelo 1: MCO, usando las observaciones 1980-2012 (T = 33)
Variable dependiente: y

             Coeficiente   Desv. Típica   Estadístico t   Valor p 
  
  const      3,69240e+11   1,22748e+10        30,08       1,67e-24 ***
  time       1,92154e+10   6,29962e+08        30,50       1,10e-24 ***

Media de la vble. dep.  6,96e+11   D.T. de la vble. dep.   1,89e+11
Suma de cuad. residuos  3,68e+22   D.T. de la regresión    3,45e+10
R-cuadrado              0,967755   R-cuadrado corregido    0,966715
F(1, 31)                930,3962   Valor p (de F)          1,10e-24
Log-verosimilitud      -846,4730   Criterio de Akaike      1696,946
Criterio de Schwarz     1699,939   Crit. de Hannan-Quinn   1697,953
rho                     0,895967   Durbin-Watson           0,241592
\end{lstlisting}
\inputencoding{utf8}

Además se puede dibujar una gráfica que recoja los valores originales
de la serie, los ajustados y los residuos. 

\inputencoding{latin9}\begin{lstlisting}[basicstyle={\scriptsize\ttfamily}]
? series yhat = $yhat
? series resid = $uhat
? gnuplot y yhat resid --time-series --with-lines
\end{lstlisting}
\inputencoding{utf8}

\begin{figure}[H]
\centering{}\includegraphics{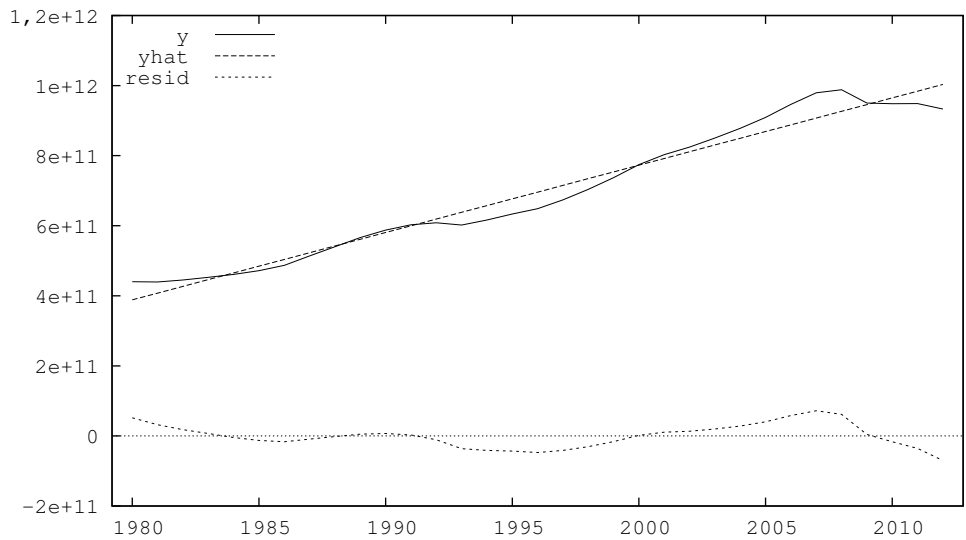}\caption{gráfico de la regresión auxiliar de $y$}
\end{figure}

El nivel de significación o la bondad del ajuste reflejan inequívocamente
la existencia de esa tendencia lineal, determinista.

A continuación se representa la variable $u$ junto a su primera diferencia. 

\inputencoding{latin9}\begin{lstlisting}[basicstyle={\scriptsize\ttfamily}]
? scatters u d_u
\end{lstlisting}
\inputencoding{utf8}

\begin{figure}[H]
\centering{}\includegraphics{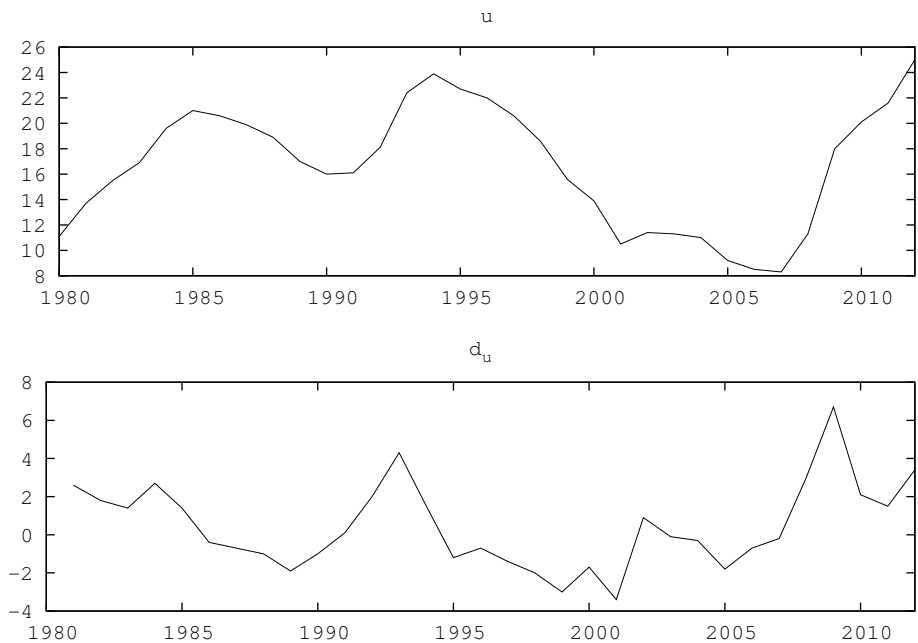}\caption{variable del desempleo a nivel y primera diferencia}
\end{figure}

En este caso no parece haber tendencia determinista. Se vuelve a utilizar
el procedimiento anterior para comprobarlo

\inputencoding{latin9}\begin{lstlisting}[basicstyle={\scriptsize\ttfamily}]
? ols u const time
\end{lstlisting}
\inputencoding{utf8}

\inputencoding{latin9}\begin{lstlisting}[basicstyle={\scriptsize\ttfamily}]
Modelo 2: MCO, usando las observaciones 1980-2012 (T = 33)
Variable dependiente: u

             Coeficiente   Desv. Típica   Estadístico t   Valor p 
  ----------------------------------------------------------------
  const      18,0712        1,71147          10,56        8,61e-12 ***
  time       -0,0820856     0,0878350        -0,9345      0,3572  

Media de la vble. dep.  16,67576   D.T. de la vble. dep.   4,794986
Suma de cuad. residuos  715,5804   D.T. de la regresión    4,804502
R-cuadrado              0,027401   R-cuadrado corregido   -0,003973
F(1, 31)                0,873370   Valor p (de F)          0,357247
Log-verosimilitud      -97,58865   Criterio de Akaike      199,1773
Criterio de Schwarz     202,1703   Crit. de Hannan-Quinn   200,1844
rho                     0,905713   Durbin-Watson           0,227574
\end{lstlisting}
\inputencoding{utf8}

\begin{figure}[H]
\centering{}\includegraphics{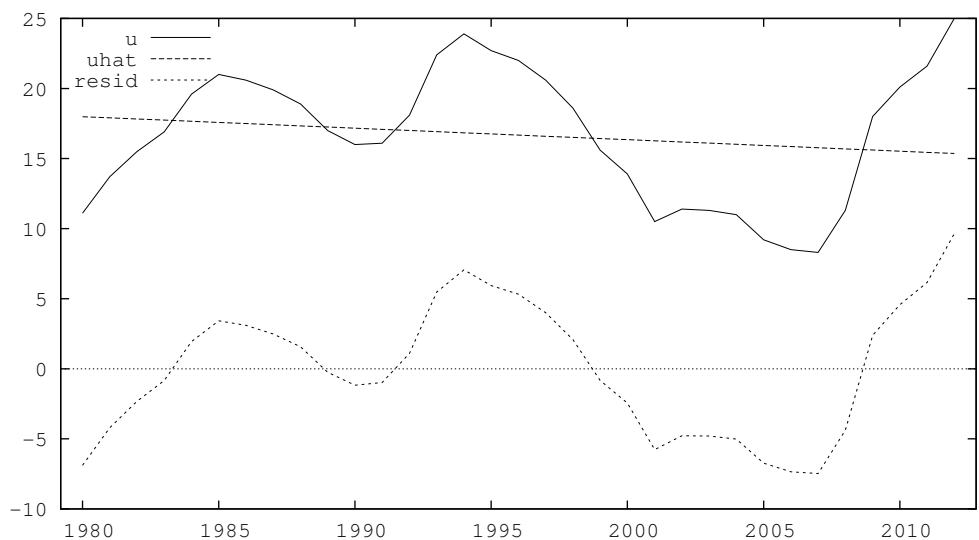}\caption{gráfico de la regresión auxiliar de $u$}
\label{3u}
\end{figure}

El p-valor es tan alto que no es posible rechazar la hipótesis nula
de no significación, con lo que la variable temporal no es relevante
en esta regresión y no habrá que tenerla en cuenta a la hora de realizar
el test aumentado de Dickey y Fuller. La figura \ref{3u} refleja
esa ausencia de tendencia determinista. Sin embargo, parece haber
tendencia estocástica en la volatilidad del gráfico de residuos. \inputencoding{latin9}
\begin{lstlisting}[basicstyle={\scriptsize\ttfamily}]
? series uhat = $yhat
? series resid = $uhat
? gnuplot u uhat resid --time-series --with-lines
\end{lstlisting}
\inputencoding{utf8}

Un tipo concreto de gráfico que resulta de gran utilidad es el correlograma.
En este, se representan las funciones de autocorrelación total y parcial
para estudiar la posible existencia de ergodicidad o esquemas autorregresivos.

\inputencoding{latin9}\begin{lstlisting}[basicstyle={\scriptsize\ttfamily}]
? corrgm y 32
? corrgm d_y 32
\end{lstlisting}
\inputencoding{utf8}

\begin{figure}[H]
\begin{centering}
\includegraphics{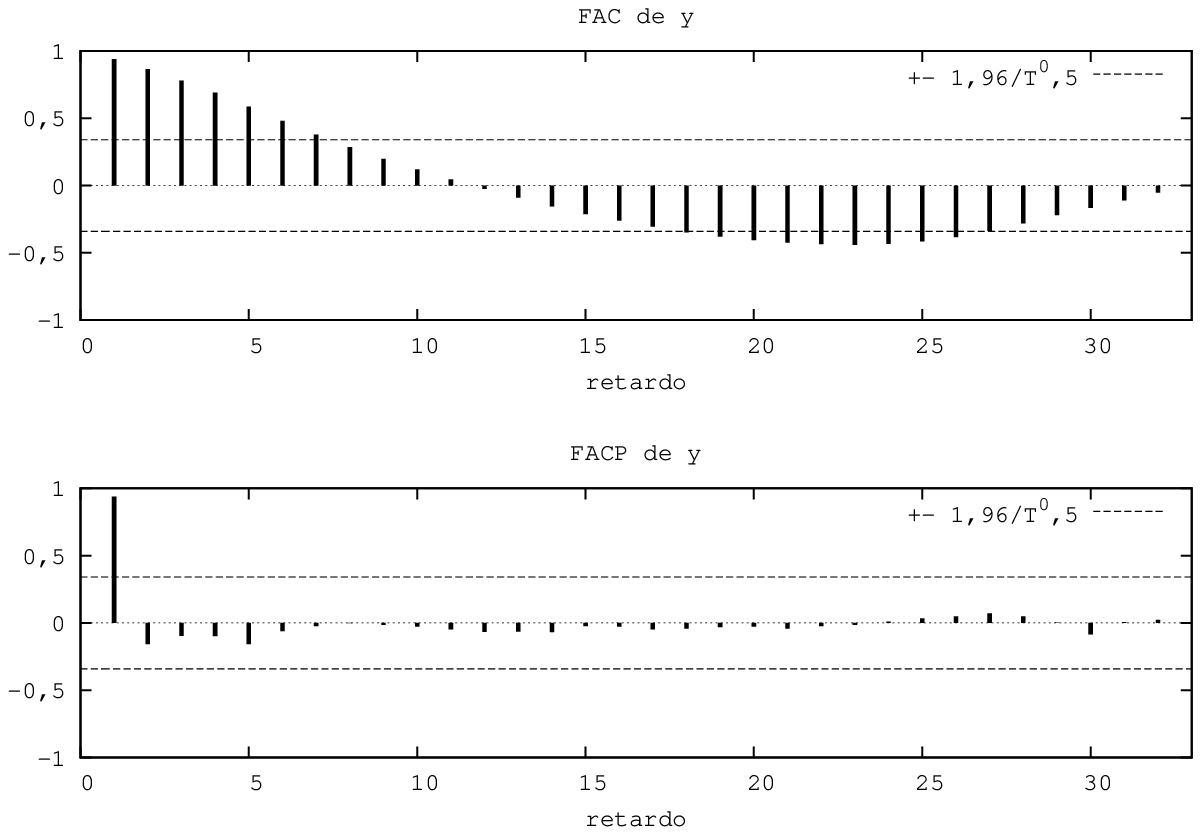}
\par\end{centering}

\centering{}\includegraphics{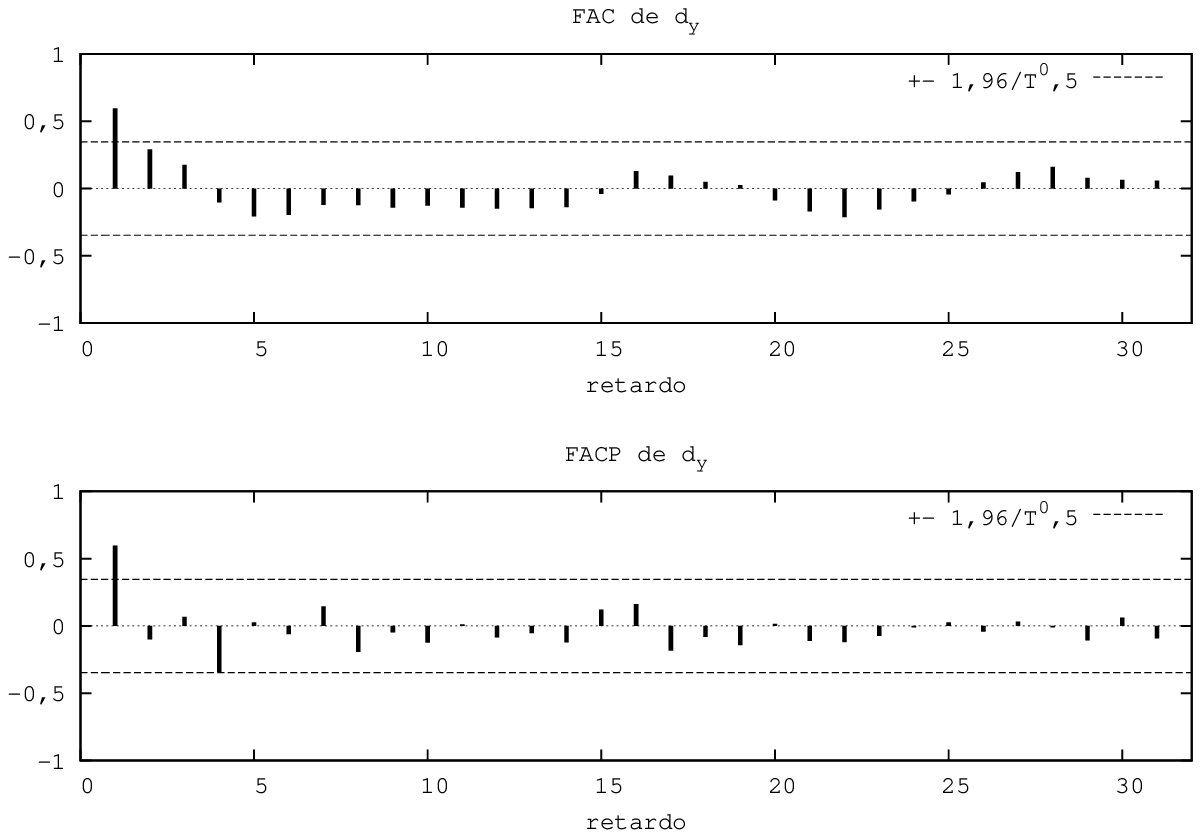}\caption{correlogramas de la serie y}
\end{figure}

El correlograma de la serie $y$ no tiende rápidamente hacia cero
por lo que no parece que haya ergodicidad.

\inputencoding{latin9}\begin{lstlisting}[basicstyle={\scriptsize\ttfamily}]
? corrgm u 32
? corrgm d_u 32
\end{lstlisting}
\inputencoding{utf8}

\begin{figure}[H]
\begin{centering}
\includegraphics{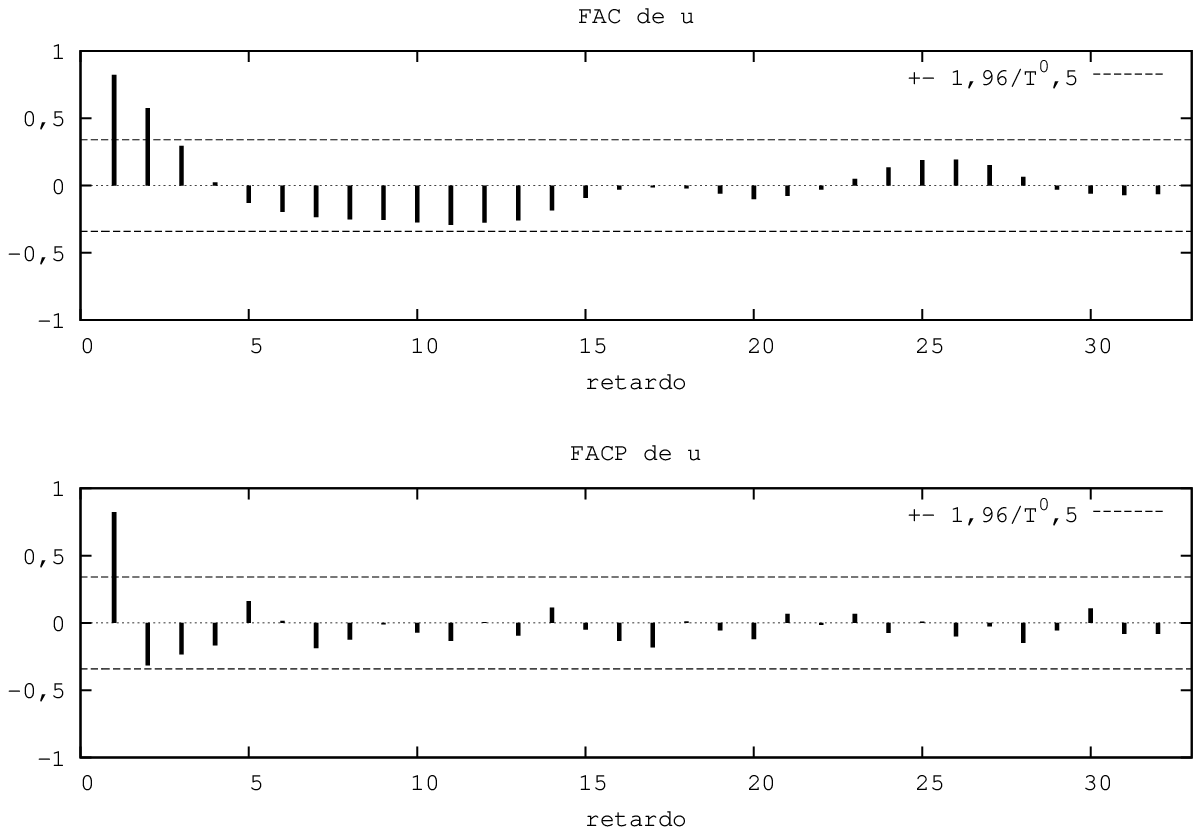}
\par\end{centering}

\centering{}\includegraphics{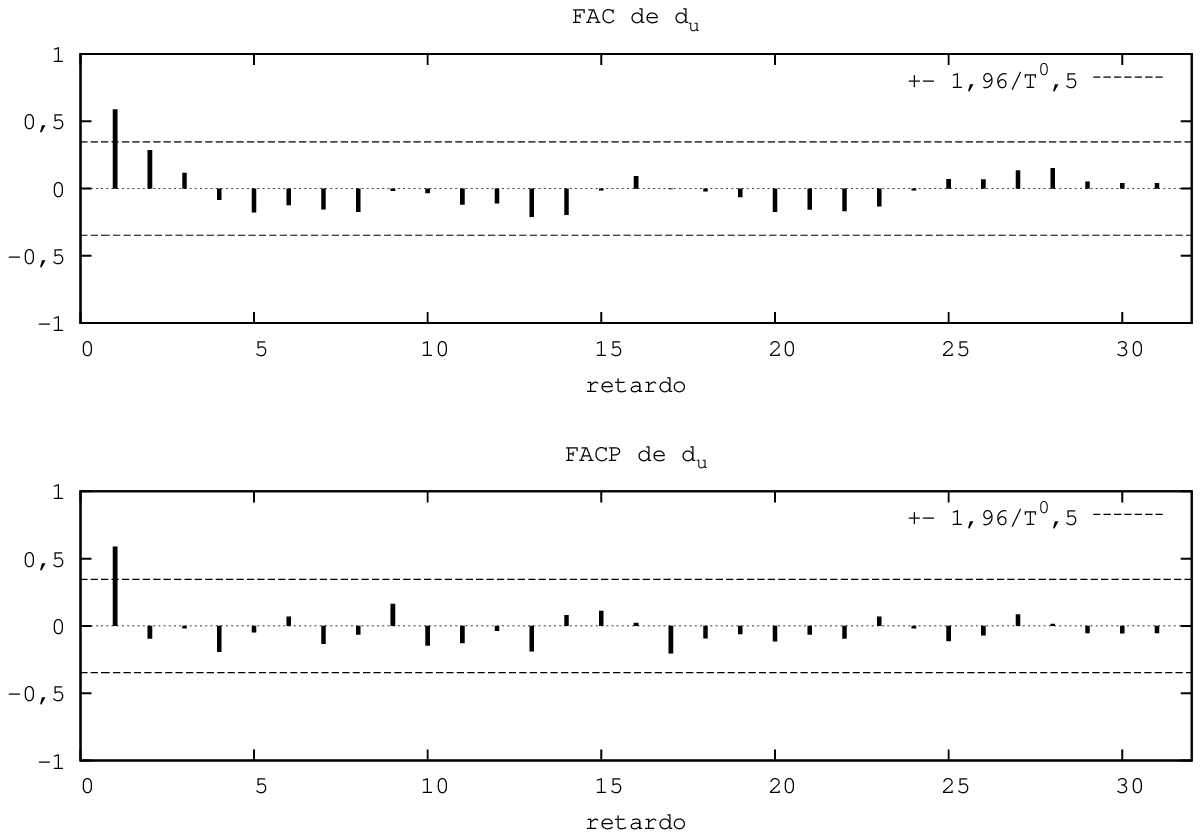}\caption{correlogramas de la serie u}
\end{figure}

Al contrario que en la anterior, la serie del desempleo sí tiende
a cero. Para completar el análisis de estacionariedad se lleva a cabo
el test aumentado de Dickey y Fuller del que ya se ha hablado en la
sección \ref{sub:adf}.

\subsubsection{ADF para la serie del PIB}

El comando específico para esta serie es el siguiente

\inputencoding{latin9}\begin{lstlisting}[basicstyle={\scriptsize\ttfamily}]
? adf 9 y --ct --test-down=MAIC
\end{lstlisting}
\inputencoding{utf8}

Proporciona la siguiente salida:
\begin{lyxcode}
\inputencoding{latin9}\begin{lstlisting}[basicstyle={\scriptsize\ttfamily},breaklines=true,extendedchars=true,tabsize=4]
MIC = 3954,46 for k = 09
MIC = 4050,33 for k = 08
MIC = 4198,62 for k = 07
MIC = 2481,89 for k = 06
MIC = 1796,71 for k = 05
MIC = 1374,59 for k = 04
MIC = 1376,44 for k = 03
MIC = 323,958 for k = 02
MIC = 243,925 for k = 01
   con constante y tendencia 
   modelo: (1-L)y = b0 + b1*t + (a-1)*y(-1) + ... + e
   Coef. de autocorrelación de primer orden de e: -0,028
   valor estimado de (a - 1): -0,14595
   Estadístico de contraste: tau_ct(1) = -1,93959
   valor p asintótico 0,6335

Regresión aumentada de Dickey-Fuller
MCO, usando las observaciones 1982-2012 (T = 31)
Variable dependiente: d_y

             Coeficiente    Desv. Típica   Estadístico t   Valor p   
  const       5,79883e+10   2,54313e+10        2,280       0,0307   **
  y_1        -0,145950      0,0752476         -1,940       0,6335  
  d_y_1       0,739899      0,145273           5,093       2,37e-05 ***
  time        2,63608e+09   1,52269e+09        1,731       0,0948   *

  AIC: 1531,79   BIC: 1537,52   HQC: 1533,66
\end{lstlisting}
\inputencoding{utf8}
\end{lyxcode}
Siguiendo el método modificado de Akaike\footnote{La regla de decisión para este criterio es $MIC_{1}\succ MIC_{2}\leftrightarrow MIC_{1}<MIC_{2}$},
el programa encuentra que el orden de rezagos óptimo es 1. Gretl calcula
el estadístico de contraste $\tau$, que es aquel que Dickey y Fuller
tabularon. Se puede hacer el contraste de hipótesis con este valor,
aunque resulta más cómodo utilizar el p-valor que ha sido calculado
correctamente, utilizando las tablas específicas del estadístico.
La hipótesis nula no puede ser rechazada para un nivel de significación
del 5\% por lo que la serie $y$ es $I(1)$.

\subsubsection{ADF para la serie del desempleo}

\inputencoding{latin9}\begin{lstlisting}[basicstyle={\scriptsize\ttfamily}]
? adf 9 u --c --test-down=MAIC
\end{lstlisting}
\inputencoding{utf8}
\begin{lyxcode}
\inputencoding{latin9}\begin{lstlisting}[basicstyle={\scriptsize\ttfamily},breaklines=true,tabsize=4]
MIC = 76,1637 for k = 09
MIC = 20,9255 for k = 08
MIC = 18,5763 for k = 07
MIC = 18,0821 for k = 06
MIC = 11,2694 for k = 05
MIC = 9,43562 for k = 04
MIC = 7,69848 for k = 03
MIC = 5,41725 for k = 02
MIC = 4,70722 for k = 01

Contraste aumentado de Dickey-Fuller para u
incluyendo un retardo de (1-L)u
(el máximo fue 9, el criterio AIC modificado)
tamaño muestral 31
hipótesis nula de raíz unitaria: a = 1

   contraste con constante 
   modelo: (1-L)y = b0 + (a-1)*y(-1) + ... + e
   Coef. de autocorrelación de primer orden de e: -0,046
   valor estimado de (a - 1): -0,150875
   Estadístico de contraste: tau_c(1) = -2,19578
   valor p asintótico 0,208

Regresión aumentada de Dickey-Fuller
MCO, usando las observaciones 1982-2012 (T = 31)
Variable dependiente: d_u

             Coeficiente   Desv. Típica   Estadístico t   Valor p 
  const        2,63039      1,16892           2,250       0,0325   **
  u_1         -0,150875     0,0687113        -2,196       0,2080  
  d_u_1        0,698815     0,143688          4,863       4,03e-05 ***

  AIC: 123,072   BIC: 127,374   HQC: 124,475
\end{lstlisting}
\inputencoding{utf8}
\end{lyxcode}
El test ADF para la serie del desempleo revela que la hipótesis nula
tampoco puede ser rechazada para un nivel de significación del 5\%
por lo que la serie $u$ es $I(1)$ al igual que la serie $y$.

\subsubsection{Contraste de cointegración}

Una vez se ha comprobado que las series son integradas de primer orden,
el análisis pasa al contraste de cointegración de Engle y Granger.
Se estima una regresión donde la tendencia temporal y la serie $y$
expliquen el comportamiento de la variable dependiente. La inclusión
de la variable temporal se produce porque la serie $y$ presentaba
tendencia determinista, como se comprobó en el apartado \ref{sub:Series-temporales}. 

Se ejecuta el siguiente comando:

\inputencoding{latin9}\begin{lstlisting}[basicstyle={\scriptsize\ttfamily}]
? coint 9 u y --test-down --skip-df --ct
\end{lstlisting}
\inputencoding{utf8}

La salida muestra que el residuo de la regresión cointegrante no es
estacionario sino integrado de orden uno. Por lo que se descarta la
existencia de una relación cointegrante que aseguraría un equilibrio
a largo plazo.
\begin{lyxcode}
\inputencoding{latin9}\begin{lstlisting}[basicstyle={\scriptsize\ttfamily},breaklines=true,tabsize=4]
Etapa 1: regresión cointegrante

Regresión cointegrante - 
MCO, usando las observaciones 1980-2012 (T = 33)
Variable dependiente: u

             Coeficiente    Desv. Típica   Estadístico t   Valor p 
  -----------------------------------------------------------------
  const      65,6535        3,64980            17,99       1,28e-17 ***
  y          -1,28866e-10   9,71955e-12       -13,26       4,44e-14 ***
  time        2,39411       0,189851           12,61       1,60e-13 ***

Media de la vble. dep.  16,67576   D.T. de la vble. dep.   4,794986
Suma de cuad. residuos  104,3197   D.T. de la regresión    1,864758
R-cuadrado              0,858211   R-cuadrado corregido    0,848759
Log-verosimilitud      -65,81569   Criterio de Akaike      137,6314
Criterio de Schwarz     142,1209   Crit. de Hannan-Quinn   139,1420
rho                     0,854909   Durbin-Watson           0,292045

Etapa 2: contrastando la existencia de una raíz unitaria en uhat

Contraste aumentado de Dickey-Fuller para uhat
incluyendo un retardo de (1-L)uhat
(el máximo fue 9, el criterio AIC modificado)
tamaño muestral 31
hipótesis nula de raíz unitaria: a = 1

   modelo: (1-L)y = (a-1)*y(-1) + ... + e
   Coef. de autocorrelación de primer orden de e: -0,082
   valor estimado de (a - 1): -0,193923
   Estadístico de contraste: tau_ct(2) = -2,01769
   valor p asintótico 0,7643

Hay evidencia de una relación cointegrante si:
(a) La hipótesis de existencia de raíz unitaria no se rechaza para las variables individuales.
(b) La hipótesis de existencia de raíz unitaria se rechaza para los residuos (uhat) de la regresión cointegrante.
\end{lstlisting}
\inputencoding{utf8}
\end{lyxcode}

\subsubsection{Modelo final}

Finalmente se plantea un modelo autorregresivo en diferencias. Este
modelo se asemeja a la ecuación generalizada de Okun I propuesta por
Belmonte y Polo de la que se hacía referencia en la subsección \ref{sub:okun generalizado}.
Al contrario que en esta, Belmonte y Polo toman logaritmos en la diferencia
de la producción. Sin embargo, en la práctica, la mejora en la bondad
del ajuste es difícilmente justificable.

\begin{equation}
\Delta u_{t}=\beta_{0}+\sum_{1}^{P}{\beta_{up}\Delta u_{t-p}}+\sum_{0}^{Q}{\beta_{yp}\Delta y_{t-q}}+\varepsilon_{t}
\end{equation}

En Gretl, la estimación de este último modelo pasa por generar las
primeras diferencias de las variables y posteriormente, realizar sucesivas
estimaciones de mínimos cuadrados ordinarios para seleccionar el órden
máximo de retardos. 

\inputencoding{latin9}\begin{lstlisting}[basicstyle={\scriptsize\ttfamily}]
? diff u
? diff y
? ols d_u const d_y
? ols d_u const d_y(0 to -1)
? ols d_u const d_y d_u(-1)
? ols d_u const d_y(0 to -1) d_u(-1)
\end{lstlisting}
\inputencoding{utf8}

Donde \texttt{d\_{[}variable{]}} es el token reservado en Gretl para
referirse a las primeras diferencias. Para añadir a un modelo de estimación
MCO los rezagos de las variables, se hace uso de la sintaxis \texttt{{[}variable{]}(0
to -{[}máximo retardo{]})}.

El cuadro \ref{cuadrofinal} resume los principales estadísticos{\scriptsize{}}\footnote{{\scriptsize{}El apartado de autocorrelación se refiere al P-valor
del test Breusch-Godfrey, con un retardo, calculado a partir del estadístico
$TR^{2}$.}} a tener en cuenta para seleccionar el orden de los retardos. Los
resultados completos de las regresiones y sus correspondientes tests
de autocorrelación pueden consultarse en el anexo. 

\begin{table}[H]
\centering{}{\scriptsize{}}%
\begin{tabular}{|c|c|c|c|c|}
\hline 
{\scriptsize{}Ecuación} & {\scriptsize{}Autocorrelación} & {\scriptsize{}$\bar{R}^{2}$} & {\scriptsize{}AIC} & {\scriptsize{}Schwarz}\tabularnewline
\hline 
\hline 
{\scriptsize{}$\Delta u_{t}=\beta_{0}+\beta_{1}\Delta y_{t}+\varepsilon_{t}$} & {\scriptsize{}0,196} & {\scriptsize{}0,799231} & {\scriptsize{}92,72732} & {\scriptsize{}95,65879}\tabularnewline
\hline 
{\scriptsize{}$\Delta u_{t}=\beta_{0}+\beta_{y0}\Delta y_{t}+\beta_{y1}\Delta y_{t-1}+\varepsilon_{t}$} & {\scriptsize{}0,26} & {\scriptsize{}0,786666} & {\scriptsize{}92,71014} & {\scriptsize{}97,01210}\tabularnewline
\hline 
{\scriptsize{}$\Delta u_{t}=\beta_{0}+\beta_{u1}\Delta u_{t-1}+\beta_{y0}\Delta y_{t}+\varepsilon_{t}$} & {\scriptsize{}0,19} & {\scriptsize{}0,786473} & {\scriptsize{}92,73821} & {\scriptsize{}97,04017}\tabularnewline
\hline 
{\scriptsize{}$\Delta u_{t}=\beta_{0}+\beta_{u1}\Delta u_{t-1}+\beta_{y0}\Delta y_{t}+\beta_{y1}\Delta y_{t-1}+\varepsilon_{t}$} & {\scriptsize{}0,331} & {\scriptsize{}0,789876} & {\scriptsize{}93,11284} & {\scriptsize{}98,84878}\tabularnewline
\hline 
\end{tabular}\caption{Selección del orden máximo de retardos}
\label{cuadrofinal}
\end{table}

Se puede concluir que el modelo más sencillo es el que ofrece mejores
resultados. \inputencoding{latin9}
\begin{lstlisting}[basicstyle={\scriptsize\ttfamily},breaklines=true,extendedchars=true,tabsize=4]
Modelo Final: MCO, usando las observaciones 1981-2012 (T = 32)
Variable dependiente: d_u

             Coeficiente    Desv. Típica   Estadístico t   Valor p 
  -----------------------------------------------------------------
  const       2,30565       0,243676            9,462      1,63e-10 ***
  d_y        -1,21453e-10   1,08890e-11       -11,15       3,39e-12 ***

Media de la vble. dep.  0,434375   D.T. de la vble. dep.   2,231101
Suma de cuad. residuos  29,98171   D.T. de la regresión    0,999695
R-cuadrado              0,805707   R-cuadrado corregido    0,799231
F(1, 30)                124,4063   Valor p (de F)          3,39e-12
Log-verosimilitud      -44,36366   Criterio de Akaike      92,72732
Criterio de Schwarz     95,65879   Crit. de Hannan-Quinn   93,69902
rho                     0,225048   Durbin-Watson           1,536587
\end{lstlisting}
\inputencoding{utf8}

\begin{figure}[H]
\centering{}\includegraphics{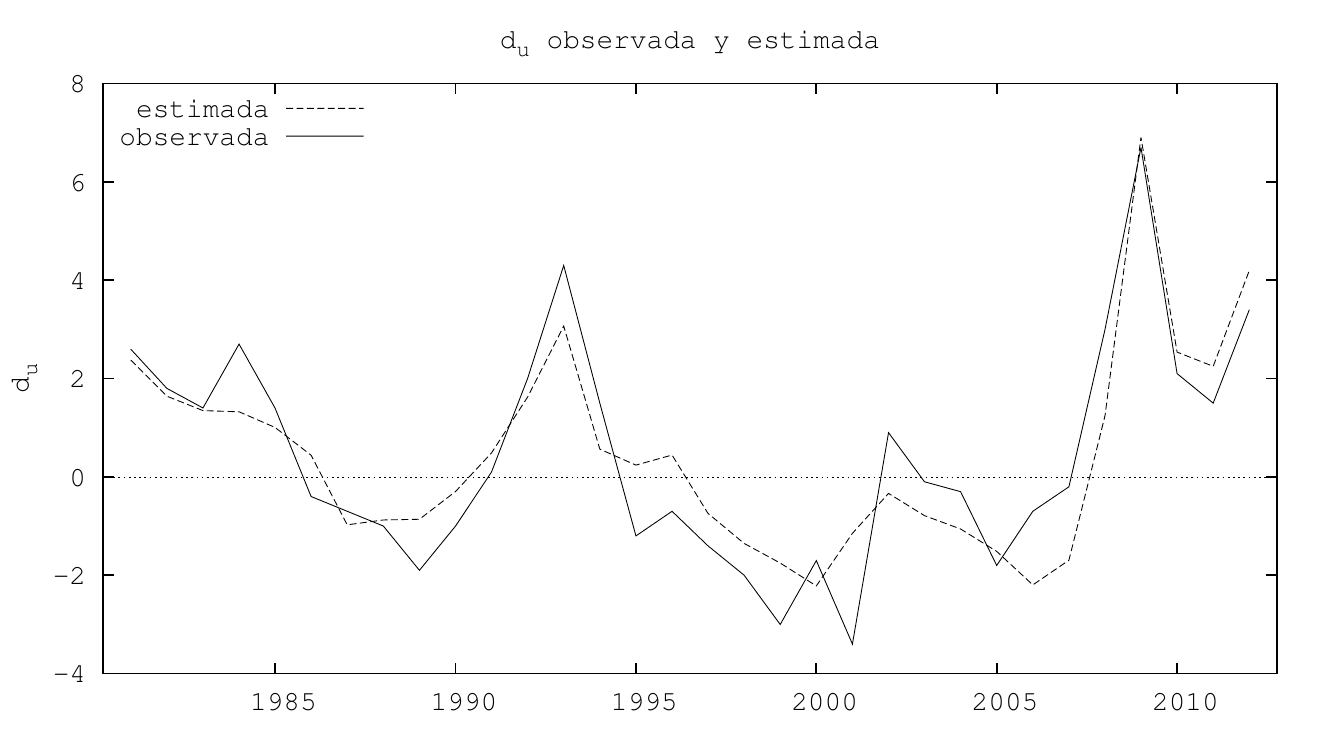}\\
\caption{datos reales vs estimados}
\label{realvsobservada}
\end{figure}

Un análisis en profundidad del modelo final, revela la ausencia de
variables no significativas individualmente, a diferencia de lo que
ocurre en los modelos con retardos. El valor de $\bar{R}^{2}$ se
utiliza como una medida de la bondad del ajuste, es decir, para medir
la parte de la varianza de la variable dependiente explicada a través
de los regresores, ajustada al número de estos. Para tratarse de datos
reales, un 80\% es un valor bastante aceptable y deja margen para
la utilización de este modelo como herramienta de análisis de políticas
de empleo. La figura \ref{realvsobservada} compara el valor estimado
de la variable dependiente con el real.

\begin{figure}[H]
\begin{centering}
\includegraphics{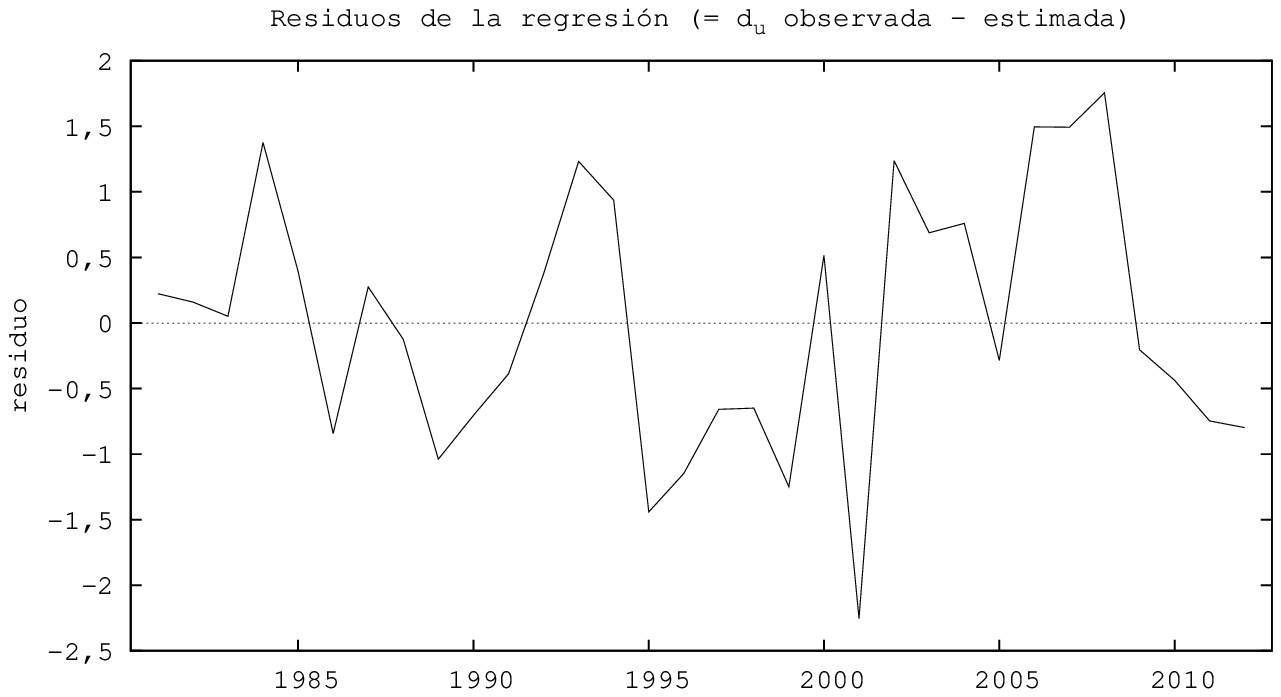}\\
\includegraphics{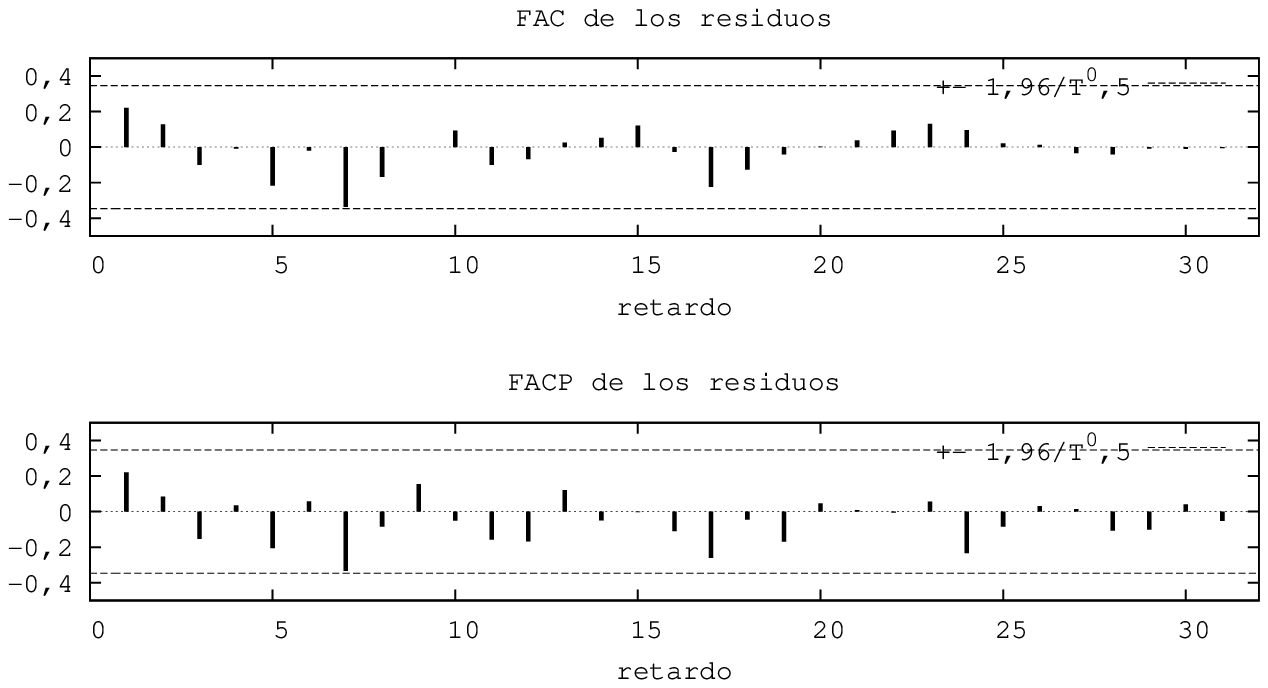}\\
\includegraphics{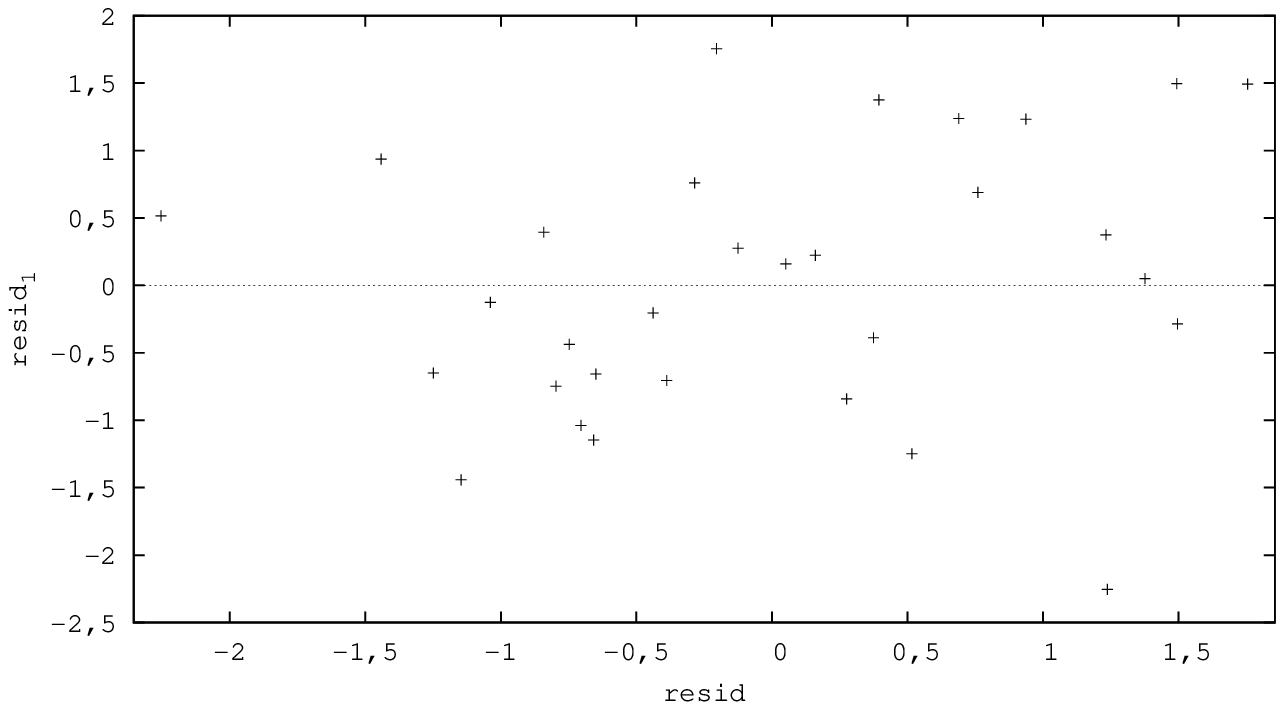}
\par\end{centering}

\centering{}\caption{gráficos de residuos}
\label{residuosfnal}
\end{figure}

La figura \ref{residuosfnal} permite estudiar en profundidad la posible
existencia de autocorrelación. Para ello se muestra el gráfico de
residuos respecto al tiempo, el correlograma y la nube de puntos que
resulta de comparar el valor del residuo con el de su retardo. 

Todo ello indica que el residuo es un ruido blanco, caracterizado
por la ausencia de autocorrelación. Para precisar aún más, se ejecuta
el test de Breusch-Godfrey con uno y dos retardos. Los resultados
reafirman la evidente ausencia de autocorrelación. \inputencoding{latin9}
\begin{lstlisting}[basicstyle={\scriptsize\ttfamily}]
? modtest --autocorr 2
\end{lstlisting}
\inputencoding{utf8}\inputencoding{latin9}\begin{lstlisting}[basicstyle={\scriptsize\ttfamily}]
Contraste Breusch-Godfrey de autocorrelación hasta el orden 2
MCO, usando las observaciones 1981-2012 (T = 32)
Variable dependiente: uhat

             Coeficiente    Desv. Típica   Estadístico t   Valor p
  ----------------------------------------------------------------
  const      -0,0868861     0,254245          -0,3417      0,7351 
  d_y         4,92629e-12   1,16092e-11        0,4243      0,6746 
  uhat_1      0,221026      0,192015           1,151       0,2594 
  uhat_2      0,113519      0,197961           0,5734      0,5709 

  R-cuadrado = 0,063296

Estadístico de contraste: LMF = 0,946019,
con valor p  = P(F(2,28) > 0,946019) = 0,4

Estadístico alternativo: TR^2 = 2,025463,
con valor p  = P(Chi-cuadrado(2) > 2,02546) = 0,363

Ljung-Box Q' = 2,30358,
con valor p  = P(Chi-cuadrado(2) > 2,30358) = 0,316
\end{lstlisting}
\inputencoding{utf8}

\inputencoding{latin9}\begin{lstlisting}[basicstyle={\scriptsize\ttfamily}]
? modtest --autocorr 1
\end{lstlisting}
\inputencoding{utf8}

\inputencoding{latin9}\begin{lstlisting}[basicstyle={\scriptsize\ttfamily}]
Contraste Breusch-Godfrey de autocorrelación de primer orden
MCO, usando las observaciones 1981-2012 (T = 32)
Variable dependiente: uhat

             Coeficiente    Desv. Típica   Estadístico t   Valor p
  ----------------------------------------------------------------
  const      -0,0548290     0,245136          -0,2237      0,8246 
  d_y         3,17485e-12   1,10699e-11        0,2868      0,7763 
  uhat_1      0,237403      0,187669           1,265       0,2159 

  R-cuadrado = 0,052295

Estadístico de contraste: LMF = 1,600240,
con valor p  = P(F(1,29) > 1,60024) = 0,216

Estadístico alternativo: TR^2 = 1,673441,
con valor p  = P(Chi-cuadrado(1) > 1,67344) = 0,196

Ljung-Box Q' = 1,703,
con valor p  = P(Chi-cuadrado(1) > 1,703) = 0,192
\end{lstlisting}
\inputencoding{utf8}

Pero todo es susceptible de mejorar. 

Uno podría pensar que la crisis económica ha producido un shock estacional
que afectaría a la bondad del modelo. Esto se podría solucionar planteando
una ecuación con variables ficticias que tomaran distintos valores
antes y después de 2007 o en los periodos en los que hubiera crecimiento/decrecimiento. 

Otra posible vía de investigación sería el planteamiento del método
de estimación por Variables Instrumentales, si la variable de la producción
fuera endógena y estuviese contemporáneamente correlacionada con la
perturbación.

También puede surgir una duda en la identificación de la variable
del desempleo como endógena, explicada por la producción. Esta diferenciación
parece arbitraria y podría superarse fácilmente con modelos de ecuaciones
simultáneas o incluso con un modelo VAR en el que ambas variables
fueran simultáneamente dependientes y explicativas, pero esto es algo
que, como dijo Sraffa\footnote{Sraffa, P. (1965). \textit{Producción de mercancías por medio de mercancías:
Preludio a una crítica de la teoría económica.} Oikos-tau.}: \textquotedbl{}...se podrá intentar más tarde, bien sea por el autor
o por alguien más joven y mejor equipado para la tarea.\textquotedbl{}

\section{Conclusiones}

Gretl ha demostrado ser una herramienta econométrica que no sólo está
a la altura del resto de soluciones comerciales sino que en muchos
aspectos supera a estas. Su potencia, sencillez y precisión proporcionan
una forma simple y rápida de elaborar modelos econométricos de series
temporales. 

En el análisis de la Ley de Okun se ha comprobado que las series de
paro y PIB son integradas de primer orden y no cointegrantes, es decir,
no hay un equilibrio a largo plazo entre las series que se han analizado.
La regresión cointegrante del test de Engle y Granger es espuria,
por lo que es incorrecto plantear un modelo MCE, ya que el término
de corrección del error sería no estacionario. 

A partir de los resultados anteriores, se ha analizado la relación
a corto plazo existente entre las variables, aplicando métodos econométricos
de selección de modelos. Todo ello ha permitido la elaboración de
un modelo concreto de la Ley de Okun para España durante el periodo
1980-2012.

\section{Anexos }

\begin{table}[H]
\begin{centering}
\begin{tabular}{|c|c|c|}
\hline 
t & u & y\tabularnewline
\hline 
\hline 
1980 & 11,1000003815 & 440112233243,992\tabularnewline
\hline 
1981 & 13,6999998093 & 439529222656,372\tabularnewline
\hline 
1982 & 15,5 & 445007786168,319\tabularnewline
\hline 
1983 & 16,8999996185 & 452884939144,963\tabularnewline
\hline 
1984 & 19,6000003815 & 460967520169,85\tabularnewline
\hline 
1985 & 21 & 471668585471,638\tabularnewline
\hline 
1986 & 20,6000003815 & 487013482004,836\tabularnewline
\hline 
1987 & 19,8999996185 & 514028717025,194\tabularnewline
\hline 
1988 & 18,8999996185 & 540215006470,599\tabularnewline
\hline 
1989 & 17 & 566291348509,902\tabularnewline
\hline 
1990 & 16 & 587705052524,646\tabularnewline
\hline 
1991 & 16,1000003815 & 602668026489,28\tabularnewline
\hline 
1992 & 18,1000003815 & 608268110818,497\tabularnewline
\hline 
1993 & 22,3999996185 & 601993875288,707\tabularnewline
\hline 
1994 & 23,8999996185 & 616340565108,611\tabularnewline
\hline 
1995 & 22,7000007629 & 633336119409,734\tabularnewline
\hline 
1996 & 22 & 648643931519,84\tabularnewline
\hline 
1997 & 20,6000003815 & 673737980443,29\tabularnewline
\hline 
1998 & 18,6000003815 & 703841675998,072\tabularnewline
\hline 
1999 & 15,6000003815 & 737245561261,189\tabularnewline
\hline 
2000 & 13,8999996185 & 774475100606,667\tabularnewline
\hline 
2001 & 10,5 & 802893926680,075\tabularnewline
\hline 
2002 & 11,3999996185 & 824653816873,566\tabularnewline
\hline 
2003 & 11,3000001907 & 850131014921,666\tabularnewline
\hline 
2004 & 11 & 877839344182,006\tabularnewline
\hline 
2005 & 9,1999998093 & 909298000000\tabularnewline
\hline 
2006 & 8,5 & 946363000000\tabularnewline
\hline 
2007 & 8,3000001907 & 979288716284,459\tabularnewline
\hline 
2008 & 11,3000001907 & 988020999700,411\tabularnewline
\hline 
2009 & 18 & 950156434697,386\tabularnewline
\hline 
2010 & 20,1000003815 & 948244130674,335\tabularnewline
\hline 
2011 & 21,6000003815 & 948721145663,275\tabularnewline
\hline 
2012 & 25 & 933148303065,534\tabularnewline
\hline 
\end{tabular}\caption{series del desempleo y producción en España 1980-2012}

\par\end{centering}

\end{table}

\inputencoding{latin9}\begin{lstlisting}[basicstyle={\scriptsize\ttfamily}]
Modelo 1: MCO, usando las observaciones 1981-2012 (T = 32)
Variable dependiente: d_u

             Coeficiente    Desv. Típica   Estadístico t   Valor p 
  -----------------------------------------------------------------
  const       2,30565       0,243676            9,462      1,63e-10 ***
  d_y        -1,21453e-10   1,08890e-11       -11,15       3,39e-12 ***

Media de la vble. dep.  0,434375   D.T. de la vble. dep.   2,231101
Suma de cuad. residuos  29,98171   D.T. de la regresión    0,999695
R-cuadrado              0,805707   R-cuadrado corregido    0,799231
F(1, 30)                124,4063   Valor p (de F)          3,39e-12
Log-verosimilitud      -44,36366   Criterio de Akaike      92,72732
Criterio de Schwarz     95,65879   Crit. de Hannan-Quinn   93,69902
rho                     0,225048   Durbin-Watson           1,536587


Contraste Breusch-Godfrey de autocorrelación de primer orden
MCO, usando las observaciones 1981-2012 (T = 32)
Variable dependiente: uhat

             Coeficiente    Desv. Típica   Estadístico t   Valor p
  ----------------------------------------------------------------
  const      -0,0548290     0,245136          -0,2237      0,8246 
  d_y         3,17485e-12   1,10699e-11        0,2868      0,7763 
  uhat_1      0,237403      0,187669           1,265       0,2159 

  R-cuadrado = 0,052295

Estadístico de contraste: LMF = 1,600240,
con valor p  = P(F(1,29) > 1,60024) = 0,216

Estadístico alternativo: TR^2 = 1,673441,
con valor p  = P(Chi-cuadrado(1) > 1,67344) = 0,196

Ljung-Box Q' = 1,703,
con valor p  = P(Chi-cuadrado(1) > 1,703) = 0,192
\end{lstlisting}
\inputencoding{utf8}

\inputencoding{latin9}\begin{lstlisting}[basicstyle={\scriptsize\ttfamily}]
Modelo 2: MCO, usando las observaciones 1982-2012 (T = 31)
Variable dependiente: d_u

             Coeficiente    Desv. Típica   Estadístico t   Valor p 
  -----------------------------------------------------------------
  const       2,25015       0,279344           8,055       9,02e-09 ***
  d_y        -1,24793e-10   1,48885e-11       -8,382       4,06e-09 ***
  d_y_1       6,18472e-12   1,55990e-11        0,3965      0,6948  

Media de la vble. dep.  0,364516   D.T. de la vble. dep.   2,232121
Suma de cuad. residuos  29,76137   D.T. de la regresión    1,030973
R-cuadrado              0,800889   R-cuadrado corregido    0,786666
F(2, 28)                56,31240   Valor p (de F)          1,54e-10
Log-verosimilitud      -43,35507   Criterio de Akaike      92,71014
Criterio de Schwarz     97,01210   Crit. de Hannan-Quinn   94,11247
rho                     0,197862   Durbin-Watson           1,589506


Contraste Breusch-Godfrey de autocorrelación de primer orden
MCO, usando las observaciones 1982-2012 (T = 31)
Variable dependiente: uhat

             Coeficiente    Desv. Típica   Estadístico t   Valor p
  ----------------------------------------------------------------
  const      -0,0372191     0,280729          -0,1326      0,8955 
  d_y         4,07690e-12   1,53254e-11        0,2660      0,7922 
  d_y_1      -2,01956e-12   1,56696e-11       -0,1289      0,8984 
  uhat_1      0,211577      0,197000           1,074       0,2923 

  R-cuadrado = 0,040971

Estadístico de contraste: LMF = 1,153463,
con valor p  = P(F(1,27) > 1,15346) = 0,292

Estadístico alternativo: TR^2 = 1,270087,
con valor p  = P(Chi-cuadrado(1) > 1,27009) = 0,26

Ljung-Box Q' = 1,27868,
con valor p  = P(Chi-cuadrado(1) > 1,27868) = 0,258
\end{lstlisting}
\inputencoding{utf8}

\inputencoding{latin9}\begin{lstlisting}[basicstyle={\scriptsize\ttfamily}]
Modelo 3: MCO, usando las observaciones 1982-2012 (T = 31)
Variable dependiente: d_u

             Coeficiente    Desv. Típica   Estadístico t   Valor p 
  -----------------------------------------------------------------
  const       2,21889       0,327253           6,780       2,30e-07 ***
  d_y        -1,17339e-10   1,52347e-11       -7,702       2,17e-08 ***
  d_u_1       0,0414554     0,114228           0,3629      0,7194  

Media de la vble. dep.  0,364516   D.T. de la vble. dep.   2,232121
Suma de cuad. residuos  29,78834   D.T. de la regresión    1,031440
R-cuadrado              0,800708   R-cuadrado corregido    0,786473
F(2, 28)                56,24875   Valor p (de F)          1,56e-10
Log-verosimilitud      -43,36911   Criterio de Akaike      92,73821
Criterio de Schwarz     97,04017   Crit. de Hannan-Quinn   94,14054
rho                     0,204602   h de Durbin             1,436546


Contraste Breusch-Godfrey de autocorrelación de primer orden
MCO, usando las observaciones 1982-2012 (T = 31)
Variable dependiente: uhat

             Coeficiente    Desv. Típica   Estadístico t   Valor p
  ----------------------------------------------------------------
  const       0,0632509     0,327780           0,1930      0,8484 
  d_y        -2,78479e-12   1,52403e-11       -0,1827      0,8564 
  d_u_1      -0,0743941     0,127590          -0,5831      0,5647 
  uhat_1      0,275252      0,218801           1,258       0,2192 

  R-cuadrado = 0,055368

Estadístico de contraste: LMF = 1,582569,
con valor p  = P(F(1,27) > 1,58257) = 0,219

Estadístico alternativo: TR^2 = 1,716418,
con valor p  = P(Chi-cuadrado(1) > 1,71642) = 0,19

Ljung-Box Q' = 1,3798,
con valor p  = P(Chi-cuadrado(1) > 1,3798) = 0,24
\end{lstlisting}
\inputencoding{utf8}

\inputencoding{latin9}\begin{lstlisting}[basicstyle={\scriptsize\ttfamily}]
Modelo 4: MCO, usando las observaciones 1982-2012 (T = 31)
Variable dependiente: d_u

             Coeficiente    Desv. Típica   Estadístico t   Valor p 
  -----------------------------------------------------------------
  const       1,66883       0,559969           2,980       0,0060   ***
  d_y        -1,20112e-10   1,52868e-11       -7,857       1,90e-08 ***
  d_y_1       3,22440e-11   2,67456e-11        1,206       0,2384  
  d_u_1       0,233909      0,195765           1,195       0,2425  

Media de la vble. dep.  0,364516   D.T. de la vble. dep.   2,232121
Suma de cuad. residuos  28,26673   D.T. de la regresión    1,023189
R-cuadrado              0,810888   R-cuadrado corregido    0,789876
F(3, 27)                38,59088   Valor p (de F)          6,65e-10
Log-verosimilitud      -42,55642   Criterio de Akaike      93,11284
Criterio de Schwarz     98,84878   Crit. de Hannan-Quinn   94,98261
rho                    -0,040878   Durbin-Watson           2,070563

Sin considerar la constante, el valor p más alto fue el de la variable 15 (d_u_1)


Contraste Breusch-Godfrey de autocorrelación de primer orden
MCO, usando las observaciones 1982-2012 (T = 31)
Variable dependiente: uhat

             Coeficiente    Desv. Típica   Estadístico t   Valor p
  ----------------------------------------------------------------
  const      -1,69054       1,95232           -0,8659      0,3945 
  d_y         1,69736e-12   1,54532e-11        0,1098      0,9134 
  d_y_1       8,72558e-11   1,00166e-10        0,8711      0,3917 
  d_u_1       0,720966      0,821219           0,8779      0,3880 
  uhat_1     -0,752540      0,832302          -0,9042      0,3742 

  R-cuadrado = 0,030484

Estadístico de contraste: LMF = 0,817518,
con valor p  = P(F(1,26) > 0,817518) = 0,374

Estadístico alternativo: TR^2 = 0,945019,
con valor p  = P(Chi-cuadrado(1) > 0,945019) = 0,331

Ljung-Box Q' = 0,0559567,
con valor p  = P(Chi-cuadrado(1) > 0,0559567) = 0,813
\end{lstlisting}
\inputencoding{utf8}

\newpage{}

\bibliographystyle{apalike}
\phantomsection\addcontentsline{toc}{section}{\refname}\nocite{*}
\bibliography{tfgclassic}

\newpage{}

\tableofcontents{}
\end{document}